\documentclass[aps,twocolumn,showpacs]{revtex4}
\usepackage{amsfonts}
\usepackage{amsmath}
\usepackage{amssymb}
\usepackage{graphicx}
\usepackage{color}
\DeclareMathOperator\erf{erf}
\usepackage{dsfont}
\begin{document}

\title{Brownian motion of free particles on curved surfaces}

\author{Ram\'on Casta\~{n}eda-Priego$^{(1)}$}
 \email{ramoncp@fisica.ugto.mx}

\author{Pavel Castro-Villarreal$^{(2)}$}
\email{pcastrov@unach.mx}

\author{Sendic Estrada-Jim\'enez$^{(2)}$}
\email{sestrada@unach.mx}

\author{Jos\'e Miguel M\'endez-Alcaraz$^{(3)}$}
\email{jmendez@fis.cinvestav.mx}

\affiliation{$^{(1)}$Divisi\'on de Ciencias e Ingenier\'ias, Campus Le\'on, Universidad de Guanajuato, Loma del Bosque 103, 37150 Le\'{o}n, Guanajuato, Mexico}

\affiliation{$^{(2)}$Centro de Estudios en F\'isica y Matem\'aticas B\'asicas y Aplicadas, Universidad Aut\'onoma de Chiapas, Carretera Emiliano Zapata, Km. 8, Rancho San Francisco, C. P. 29050, Tuxtla Guti\'errez, Chiapas, Mexico}

\affiliation{$^{(3)}$Departamento de F\'isica, Cinvestav, Av. IPN 2508, Col. San Pedro Zacatenco, 07360 M\'exico, D. F., Mexico}

\begin{abstract}
Brownian motion of free particles on curved surfaces is studied by means of the Langevin equation written in Riemann normal coordinates. In the diffusive regime we find the same physical behavior as the one described by the diffusion equation on curved manifolds [J. Stat. Mech. (2010) P08006]. Therefore, we use the latter in order to analytically investigate the whole diffusive dynamics in compact geometries, namely, the circle and the sphere. Our findings are corroborated by means of Brownian dynamics computer simulations based on a heuristic adaptation of the Ermak-McCammon algorithm to the Langevin equation along the curves, as well as on the standard algorithm, but for particles subjected to an external harmonic potential, deep and narrow, that possesses a ``Mexican hat" shape, whose minima define the desired surface. The short-time diffusive dynamics is found to occur on the tangential plane. Besides, at long times and compact geometries, the mean-square displacement moves towards a saturation value given only by the geometrical properties of the surface.
\end{abstract}


\maketitle

\section{Introduction}

During last decades, the interest in diffusive processes has grown tremendously because of their universality in diverse physical areas; ranging from condensed matter to elementary particle physics and gravitation \cite{Soft, quark-diffusion, Hu}. In particular, it has emerged an intense activity in the study of Brownian motion in curved manifolds motivated by problems coming from biophysics \cite{biophysics}.  For instance,  the lateral diffusion of proteins and lipids occurring inside cell membranes are interesting and complex since they determine the flux of nutrients between the cell and its exterior affecting, in consequence, the cell functionality \cite{Alberts}. From the theoretical point of view, it is difficult to describe this phenomenon because the interactions with the remaining components of the membrane and the protein finite-size effects \cite{Naji,fidel1,fidel2,Reister1}. Besides, there are also curvature contributions \cite{faraudo02} and thermal fluctuations that produce shape undulations \cite{Gustafsson} coupled to the lateral motion \cite{Reister}. And on top of that, protein diffusion is also affected by changing  membrane thickness \cite{Gov, naohisa}.  The simplest approach to study this problem is to consider the Brownian motion of a punctual particle on a frozen two-dimensional regular surface that represents the  membrane \cite{Aizenbaud, H, K, faraudo02, Holyst, Christensen, Tomoyoshi, Castro}. In this approximation, both thermal shape fluctuations and finite-size effects have not been taken into account explicitly but as an effective result reflected in the parameters of the model. As discussed below and although the results presented here are quite general for the Brownian motion on a manifold, this work is primarily motivated by the aforementioned transport phenomena. 

Although the understanding of Brownian motion was established a century ago, it is noteworthy to mention that the study of Brownian motion on manifolds started three decades ago. Since the seminal work of N. G. van Kampen \cite{Kampen}, the fundamental equations of Brownian motion on manifolds were established and the manifolds introduced, like in classical mechanics, as a result of the appropriate canonical transformations involved in the system with certain holonomic constraints.  Manifolds also appeared naturally in the dynamics of polymers in solution \cite{Fixman, Rallison, Hinch}, when the polymer is modeled by means of the theory of Brownian motion with constraints (see, e.g., \cite{Ottinger}, and more recently \cite{Morse} for a review). In this case, the number of constraints that take into account the bonds between monomers establishes the dimension of the manifold. However, in a real situation, the rigid constraints represent idealizations of stiff potentials that limit the motion in a certain spatial domain \cite{KampenandLodder}, whereas by including either thermal or statistical fluctuations the rigid constraints, in general, will no longer represent idealizations of elastic potentials \cite{KampenandLodder, Rallison, Hinch}. Nonetheless, albeit the fluctuations are present, the rigid constraints may emulate real molecular bonds as illustrative toy models and, in some cases, realistic models can be also represented through a coarse-grain or large-scale description, like in the rigid-rod and wormlike chain models \cite{Ottinger}. 

In addition, Brownian dynamics on curved manifolds becomes a natural framework to study diffusion on crystals with topological disorder, where the torsion of the manifold is crucial to quantify the degree of disorder \cite{Kleinert, Baush, Lingand Chen}. Furthermore, Smerlak has found that the Eckart's heat flux in General Relativity and the generalization of the Tolman-Ehrenfest relation to non-equilibrium stationary states, as well as gravitational corrections, can be best understood through the mean-square displacements of hypothetic particles in static isotropic curved space-times \cite{Smerlak}.

Although the diffusion equation is suited to study the Brownian motion of free particles on curved surfaces, a more complete description is provided by the Langevin equation. The latter is based on the Newton's equation of motion but including a rapidly fluctuating force, Gaussian distributed, representing the interaction among the particle and the solvent. It is well-known that in Euclidean open spaces the mean-square displacement (MSD) calculated from the Langevin equation reproduces the standard Einstein kinematical relation. In this kind of spaces, both Langevin and diffusion equations describe the same dynamical behavior at the diffusive time regime, i.e., $t\gg \tau_{B}=M/\zeta$, where $\zeta$ is the friction coefficient of the solvent, $M$ is the particle mass and $\tau_{B}$ the momentum relaxation time \cite{Dhont}. In a curved space, one might ask whether this property is preserved and, in general, what is the dependence of the dynamics on the geometry of the space.  These points have been recently addressed by M. Polettini \cite{Polettini} whom posed a Langevin equation, derived by a Gauge principle and proved that its overdamped limit corresponds to the diffusion equation in curved manifolds. We here discuss the aforementioned points, but taking the damped ($t\gg \tau_{B}$) and overdamped ($t\to\infty$) limits in the MSD and look at its behaviour as a function on the geometrical properties of the space.

In this work, we write down the Langevin equation for manifolds following the same method introduced by E. J. Hinch \cite{Hinch}. The starting point is the Newton's equation for free particles in a $d$-dimensional hypersurface $\mathbb{M}$. Free means here that particles do not interact between each other and non external force is acting on them. Nevertheless, they are restricted to move on $\mathbb{M}$. For the local momenta $p^{a}$ and local coordinates $x^{a}$, the resulting Langevin equations are,
\begin{eqnarray}
\label{LocalL-1}
\dot{p^{c}}&=&-\frac{1}{\tau_{B}}p^{c}-\frac{1}{M}\Gamma^{c}_{~ba}p^{b}p^{a}+f^{c},\nonumber\\
\dot{x}^{a}&=&g^{ab}p_{b}/M,
\label{LocalL-2}
\end{eqnarray}
where $g_{ab}$ is the Riemannian metric tensor and $\Gamma^{c}_{~ba}$ the Christoffel symbols. It turns out that these equations are the same found by Kleinert and Shabanov \cite{Kleinert}, who discussed its generalization to connections with torsion, as well as those derived in the work of M. Polettini using a Gauge principle (invariance under local rotations) \cite{Polettini}. Besides, the global version of these equations were obtained by E. J. Hinch \cite{Hinch} for the particular case of two monomers, one of them excessively massive, with one constraint. 

We also find that the MSD, up to first order in curvature, calculated from (\ref{LocalL-2}) is given by
\begin{eqnarray}
\label{MSD0-1}
\left<s^{2}(t)\right>&=&2dD_{0}\tau_{B}\left[\frac{t}{\tau_{B}}-\frac{1}{2}(e^{-2\frac{t}{\tau_{B}}}-1)  -2(1-e^{-\frac{t}{\tau_{B}}})\right]\nonumber\\
&-&\frac{2R_{g}}{3}\left(\tau_{B}D_{0}\right)^{2}\mathcal{J}\left(t/\tau_{B}\right)+\cdots,
\label{result1}
\end{eqnarray}
where the terms in the square parenthesis are found to be the standard MSD for the particle dynamics in the Euclidean $\mathbb{R}^{d}$ space and $\mathcal{J}\left(x\right)$ is a non-dimensional function that characterizes the particle dynamics coupled to the curvature (see below at appendix B for its definition). $R_{g}$ is the Ricci scalar curvature and $s$ is the geodesic distance of the general Riemannian manifold. Taking this equation, we are able to investigate the particle dynamics at different time scales: $\tau_{solvent} \ll t \ll \tau_{B}$ and for $\tau_{B} \ll t < \tau_{G}$. $\tau_{solvent}$ is a characteristic time for the positions and momenta relaxation of the solvent molecules, at which the Langevin description is not longer valid, and $\tau_{G}$ is the time scale when curvature effects become evident. It is shown that equation (\ref{MSD0-1}) reproduces the same leading curvature effects in the diffusion regime as in Ref. \cite{Castro}, which is  based on the diffusion equation on curved manifolds. 

The geometrical properties become evident at times $t$ much more longer than $\tau_{G}$. When this happens, the system reaches the thermodynamical equilibrium. In this regime, we reproduce the free-particle dynamics based on the diffusion equation on curved manifolds. The latter is explicitly compared with our computer simulations. Using the well-known result that for compact supports ($\subset\mathbb{M}$) the spectra of the Laplace-Beltrami operator, $\Delta_{g}$, is discrete. Thus, it is easy to find that the expectation value for any observable $\mathcal{O}$ in the overdamped limit is,
\begin{eqnarray}
\label{obev1}
\left<\mathcal{O}\left(x\right)\right>\approx \frac{1}{v}\int dv ~\mathcal{O}\left(x\right)+O\left(e^{-D_0 t \lambda_{1}}\right),
\end{eqnarray}
where $\lambda_{1}\sim~1/\left(D_{0}\tau_{G}\right)$ is the first non-zero  eigenvalue of $\Delta_{g}$ and $dv$ is the volume element of the Riemannian geometry \cite{Grygorian}. It is remarkable that the leading term obtained in this way allows us to determine the steady spacial density
\begin{eqnarray}
P^{*}\left(x\right)=\frac{1}{v}\sqrt{g\left(x\right)},
\end{eqnarray}
where $g=\det g_{ab}$ and $v$ the volume of $\mathbb{M}$. This density (see appendix D) is consistent with the original calculation performed by Kramers \cite{Kramers} and recently discussed in \cite{Polettini}. We explicitly analyze the dynamics of particles confined along a circle, as well as on a sphere.



We test equations (\ref{MSD0-1}) and (\ref{obev1}) by means of Brownian dynamics computer simulations based on an heuristic adaptation of the Ermak-McCammon algorithm \cite{Ermak} to the Langevin equation along curves, as well as on the standard algorithm. In the first case, which is here only applied to the circle, the particles are allowed to move in any direction with equal probability, but the geodesic distances they travel are Gaussian randomly distributed. In the second case, the particles are subjected under the action of a spring-like force field in the $2$ ($3$)-dimensional Euclidean space, where the corresponding potential, with a ``Mexican hat" shape, has its minima at the same points of the circle (sphere). In the limit case of very stiff springs, we get the same results from both numerical routes, and analytical one,  as we will see further below. We should mention that the inclusion of a spring-like potential to reproduce the holonomic constrain is a controversial issue because the agreement between theory and simulations establishes a clear example where the particles dynamics with Lagrange constraints is equivalent with that using the stiff elastic potential even in systems with fluctuations. Moreover, we should point out that this is not in contradiction with the work done by E. J. Hinch \cite{Hinch} and Kampen and Lodder \cite{KampenandLodder}. In particular, it is shown that the single canonical partition function using the stiff potential posed is the same for the single canonical partition function on the circle (sphere) in the limit of very stiff springs field, as far as the spring-like constant $\kappa$ scales with the square of temperature. 

After the Introduction, the manuscript is organized as follows. In section II we present the Langevin equation for curved manifolds, written in both global and local coordinates. In addition, we study the curvature effects on the MSD at the following time regimes: $\tau_{solvent} \ll t \ll \tau_{B}$ and $\tau_{B} \ll t < \tau_{G}$. In section III we study the particle dynamics on the geometrical regime ($t \gg \tau_{G}$) by means of the diffusion equation on curved manifolds. In section IV we explicitly compare the predictions for particles restricted to move along a circle and on a sphere with Brownian dynamics computer simulations. Finally, in section V we summarize some concluding remarks and perspectives of our work. 

\section{Langevin equation on curved manifolds}

\subsection{Global coordinates description}

We now specify the basis of  Langevin dynamics formalism following the method introduced by E. J. Hinch \cite{Hinch}. It is defined over an Euclidean hypersurface $\mathbb{M}\subset \mathbb{R}^{d+1}$, which is represented as the points ${\bf X}\in \mathbb{R}^{d+1}$ such that $\Phi\left({\bf X}\right)=0$. The Langevin equation needs to include an holonomic constraint in order to bound a point particle on $\mathbb{M}$.

Let us denote the momentum ${\bf P}\in T_{{\bf X}}(\mathbb{M})$, where $T_{{\bf X}}(\mathbb{M})$ is the tangent space at the point ${\bf X}$, i.e, the position ${\bf X}\in\mathbb{R}^{d+1}$ of the particle. From a classical mechanics point of view, the addition of the term $\lambda\Phi({\bf X})$ to the free-particle Lagrangian allows us to impose an holonomic constraint on $\mathbb{M}$. Indeed, the resulting equation of motion is $\dot{\bf P}=\lambda ~\nabla \Phi\left({\bf X}\right)$ and the required constraint is $\Phi({\bf X})=0$. We should remark that $\lambda=0$ relaxes the constraint. Then, for the Langevin equation defined on $\mathbb{M}$, we simply include the previous constraint, a friction term and a stochastic force ${\bf f}\left(t\right)$
\begin{eqnarray}
\label{LEq1}
\dot{\bf P}&=&-\zeta~ {\bf P}/M+\lambda~\nabla\Phi\left({\bf X}\right)+{\bf f}\left(t\right)\\
\label{LEq2}
{\dot{\bf X}}&=&{\bf P}/M,\\
\label{GEq3}
\Phi\left({\bf X}\right)&=&0.
\end{eqnarray}
The second term of the right-hand side of equation (\ref{LEq1})  represents the force caused by the holonomic constraint. The stochastic force  is chosen such that it satisfies the standard  fluctuation-dissipation relations 
\begin{eqnarray}
\left<f_{i}(t)\right>&=&0,\nonumber\\
\left<f_{i}(t) f_{j}(t^{\prime})\right>&=&\Omega\delta_{ij}\delta(t-t^{\prime}),
\label{fluc-dissip}
\end{eqnarray}
where $\left<\cdots\right>$ stands for the average in the {\it ensemble} of forces  Gaussian distributed over $\mathbb{R}^{d+1}$ space (see Appendix A). Remark that $\mathbb{R}^{d+1}$ is a copy of $T_{{\bf X}}(\mathbb{M})\times\mathbb{R}$, i.e., the ensemble is given by all possible configurations of forces belonging to $\mathbb{R}^{d+1}$. The stochastic forces can be treated as vector fields in one dimension in the same spirit that Zinn-Justine introduced them in \cite{Zinn-Justin}. 

The Lagrange multiplier $\lambda$ can be obtained using the constraint (\ref{GEq3}) as follows. A time derivative on this constraint implies that
\begin{eqnarray}
\label{constraint1}
\nabla\Phi\left({\bf X}\right)\cdot{\bf P}=0,
\end{eqnarray} 
where $\nabla$ represents derivations in the space $\mathbb{R}^{d+1}$. Since the momentum ${\bf P}\in T_{\bf X}(\mathbb{M})$, then $\nabla \Phi$ is normal to the tangent space. Thus, the normal vector to the surface, i.e., normal to $T_{\bf X}(\mathbb{M})$, is given by ${\bf n}=\nabla\Phi\left({\bf X}\right)/\left|\nabla\Phi\left({\bf X}\right)\right|$. Second derivative on equation (\ref{constraint1}) gives
\begin{eqnarray}
\label{constraint2}
{\bf n}\cdot{\dot{\bf P}}=-\frac{1}{M}P^{i}G_{ij}P^{j},
\end{eqnarray}
with $G_{ij}=\partial_{i}\partial_{j}\Phi/\left|\nabla\Phi\right|$. Now, we get $\lambda$ by equa\-ting (\ref{constraint2}) and the normal projection of (\ref{LEq1}). Then, $\lambda=-P^{i}G_{ij}P^{j}/M\left|\nabla\Phi\right|-{\bf n}\cdot{\bf f}/\left|\nabla\Phi\right|$.  Therefore, the Langevin equation involves a non-linear term proportional to a second power in momenta,
\begin{eqnarray}
\dot{\bf P}+\frac{1}{M}G_{ij}P^{i}P^{j}{\bf n}&=&-\frac{\zeta}{M}~{\bf P}+\mathbb{P}\left({\bf f}\left(t\right)\right),
\label{LangP}
\end{eqnarray}
and a projector, $\mathbb{P}=\mathds{1}-{\bf n}\otimes{\bf n}$, that maps a vector ${\bf v}\in T_{{\bf X}}(\mathbb{M})\times\mathbb{R}\cong \mathbb{R}^{d+1} $ into the tangent space. We point out that the $G$ matrix encodes the surface geometry. For instance,  the constraint $\Phi({\bf X})={\bf a}\cdot{\bf X}+b$ defines a plane in Euclidean space, where ${\bf a}$ is a constant vector and $b$ a real number. In this particular case, the $G$ matrix is zero and the normal  vector of the surface is constant, ${\bf n}={\bf a}/\left|{\bf a}\right|$, as it is required for a planar geometry. In the case of a sphere of radius $R$, we have $\Phi({\bf X})={\bf X}^{2}/R^{2}-1$ and the normal vector satisfies ${\bf n}={\bf X}/R$; the matrix $G$ is given by $G_{ij}=\delta_{ij}/R$.

We should remark that the way in which the constraint affects the fluctuating force is through the projector $\mathbb{P}$. In other words, although there is a distribution of forces in $\mathbb{R}^{d+1}$ for each point of the manifold, the Langevin equation (\ref{LangP}) takes into account just those forces tangent to $\mathbb{M}$ through the projector $\mathbb{P}$. It is also remarkable that for the constrained dynamics, for instance in a numerical routine,  the fluctuating forces can be implemented in the same way as it is done for the three-dimensional Euclidean spaces.   Also, one has to note that the quadratic term in the momentum is not a surprise since the left-hand side of equation (\ref{LangP}) corresponds to the ordinary kinetic term for a particle over a hypersurface. This means that the Langevin equation reduces to the geodesic equation when both the friction and the stochastic force vanish together. This will be clarified further below when we write down the equation in local coordinates. We also have to mention that this equation is a particular case of a more general equation derived first by E. J. Hinch \cite{Hinch} within the context of polymers in solution for the case of two monomers, one of them excessively massive,  with a single constraint. In addition, this global description is the natural starting point to introduce ambient interactions, where the extrinsic geometry may play a crucial role.

It is also important to mention that constraints ``are merely the result of elastic forces excerted by connecting strings or rods, or other devices by which the free motion is hindered" \cite{KampenandLodder} and it would not be the exception for integral proteins or lipids in plasma membranes. Thus, it is natural to ask whether the Lagrange constraints are idealizations of elastic potentials when Langevin-type of forces are present. To answer properly this question it would be necessary a careful analysis and it is out from the scope  of the paper. However, following the analysis by N. G. van Kampen and J. J. Lodder \cite{KampenandLodder} one can conclude that a constraint system, with Langevin-type of forces, could be the limiting case of an equivalent stiff system provided, minimally, that these Langevin rapidly fluctuating forces, as well as all the remainder external forces, act upon the particle during a short-time $\delta$ with the requirement that $\delta\gg 1/\sqrt{k}$, where $k$ is the stiffness parameter. For instance, one can choose $\tau_{solvent}$ i.e., mean collision time of the solvent molecules, for $\delta$. In general, as it is observed by E. J. Hinch \cite{Hinch}, it is necessary to introduce an extra {\it pseudo-corrective force} in order to convert the Brownian motion of a constrained system into an equivalent very stiff system. In section IV, we test equation (\ref{result1}), which is a consequence of the constrained Langevin equation, in the cases of a sphere and a circle using the Ermak-McCammon algorithm implemented by a particle immersed in stiff elastic potential. It will be proved that, in these particular cases, the constrained system is equivalent to that of very stiff potential. 

\subsection{From a global to a local coordinates description}
 
We now provide a description in local coordinates of the Langevin equation (\ref{LangP}). In local coordinates a hypersurface is parametrized by the mapping ${\bf X}:U\subset \mathbb{R}^{2}\to \mathbb{M}$, where a particular point in $\mathbb{M}$ is given by ${\bf X}\left(x^{a}\right)$, being $x^{a}$ the local coordinates ($a=1, \cdots, d$). In such coordinates,  we have $\dot{\bf X}={\bf e}_{a}\dot{x}^{a}$,  ${\bf P}={\bf e}_{a}p^{a}$, and $\mathbb{P}\left({\bf f}\right)={\bf e}_{a}f^{a}$, where $p^{a}=g^{ab}p_{b}$ is the local momentum and ${\bf e}_{a}=\partial_{a}{\bf X}$ the tangent vectors (note that $\partial_{a}\equiv\partial/\partial x^{a}$). Thus, the first derivative of the momentum is given by
\begin{eqnarray}
\dot{\bf P}&=&\frac{1}{M}\partial_{b}{\bf e}_{a}p^{a}p^{b}+{\bf e}_{a}\dot{p^{a}},
\end{eqnarray}
where $\dot{p^{a}}\equiv d(g^{ab}p_{b})/dt$. The partial derivative $\partial_{b}{\bf e}_{a}$ can be calculated using the Weingarten-Gauss equations $\partial_{a}{\bf e}_{b}=\Gamma^{c~}_{~ba}{\bf e}_{c}-K_{ba}{\bf n}$, where $K_{ab}$ are the components of the second fundamental form \cite{Spivak}. By using these equations in the momentum time derivative we obtain
\begin{eqnarray}
\dot{\bf P}=\left(\frac{1}{M}\Gamma^{c}_{~ba}p^{b}p^{a}+\dot{p^{c}}\right){\bf e}_{c}-\frac{1}{M}K_{ab}p^{a}p^{b}{\bf n}.
\label{1st}
\end{eqnarray}
The local coordinates version of the Langevin equation can be straightforwardly obtained by substituting equation (\ref{1st}) into equation (\ref{LangP}).  Hence, the tangent projection takes the form,
\begin{eqnarray}
\label{LocalL1}
\dot{p^{c}}&=&-\frac{\zeta}{M}p^{c}-\frac{1}{M}\Gamma^{c}_{~ba}p^{b}p^{a}+f^{c},\nonumber\\
\dot{x}^{a}&=&g^{ab}p_{b}/M,
\label{LocalL2}
\end{eqnarray}
while the normal projection is given by
\begin{eqnarray}
K_{ab}={\bf e}^{i}_{a}G_{ij}{\bf e}^{j}_{b}.
\label{normalR}
\end{eqnarray}
Equations in (\ref{LocalL2}) are the local version of the Langevin equation (\ref{LangP}). They are the same derived by Kleinert and Shavanov who discussed the case of manifolds with torsion, see e.g., \cite{Kleinert}. The same equations were also obtained by M. Polettini from the local rotational invariance of  Wiener increments \cite{Polettini}. This Gauge invariance is also noted in the distribution of the forces (\ref{Distribution}). As we mentioned above, the quadratic contribution in momentum is just the geodesic contribution. The normal projection (\ref{normalR}) provides a geometrical identity that allows us to derive the extrinsic curvature in terms of the $G$ matrix. This identity is not casual;  it is actually the same found at the level set formulation of differential geometry \cite{Spivak}. 

Regarding the fluctuation-dissipation relations, the stochastic forces satisfy the following properties,
\begin{eqnarray}
\left<f_{a}(t)\right>&=&0,\nonumber\\
\left<f_{a}(t)f_{b}(t^{\prime})\right>&=&\Omega\delta_{ab}\delta(t-t^{\prime}),
\label{locfluc-diss}
\end{eqnarray}
where $\delta_{ab}$ is the two-dimensional Kronecker's delta. These relations are equivalent to their global version (see Appendix A).

\subsection{Dynamics beyond a local neighborhood}

Based on equation (\ref{LocalL2}), it is clear that the particle dynamics does not depend on the extrinsic properties of the geometry. This means that the dynamics on a hypersurface can be studied in a Riemannian geometry; this is what we do from now on. We are mainly interested on the diffusion mechanisms in the weak curvature regime. Let us recall that if $V\subset \mathbb{M}$ is a local neighborhood of $\mathbb{M}$, the map ${\bf X}:U\subset{\mathbb{R}^{d}}\to V$ is a local diffeomorphism \cite{Spivak}, then $V\equiv {\bf X}(U)\cong \mathbb{R}^{d}$. This implies that in a local neighborhood, we should have the same particle dyna\-mics as found in planar spaces (see, e.g., Ref. \cite{Dhont} for the $\mathbb{R}^{3}$ case). Thus, it makes sense to study curvature effects around the Euclidean solution.

Then, we first review the particle dyna\-mics on the Euclidean geometry $\mathbb{M}=\mathbb{R}^{d}$, i.e., when the curvature is zero, and, second, we expand the Euclidean solution in order to study the leading curvature effects on the particle dyna\-mics over the surface.

\subsubsection{Euclidean geometry $S=\mathbb{R}^{d}$}

In the Euclidean geometry, both the global and local descriptions are the same; the Euclidean metric is simply $g_{ab}=\delta_{ab}$ and the Chrystoffel symbols are zero.  In this case, the Langevin dynamics formalism reduces to the well-known standard equations \cite{Dhont}
\begin{eqnarray}
\label{euclideang}
\dot{p}^{c}&=&-\frac{\zeta}{M}p^{c}+f^{c},\nonumber\\
\dot{x}^{c}&=&\frac{1}{M}p^{c},
\end{eqnarray}
and their solution can be written as \cite{Zinn-Justin,Dhont},
\begin{eqnarray}
 \label{pplanar}
 p^{c}(t)&=&p^{c}_{0}e^{-\frac{\zeta}{M}t}+\int_{0}^{t}dt^{\prime}f^{c}(t^{\prime})e^{-\frac{\zeta}{M}\left(t-t^{\prime}\right)},\nonumber\\
 x^{c}(t)&=&x^{c}_{0}+\frac{1}{M}\int_{0}^{t}dt^{\prime}p^{c}(t^{\prime}).
\end{eqnarray}
Averaging equations (\ref{pplanar}) over the ensemble of stochastic forces, one easily obtains
\begin{eqnarray}
\left<p^{c}(t)\right>&=&p^{c}_{0}e^{-\frac{\zeta}{M}t},\nonumber\\
\left<x^{c}(t)\right>&=&x^{c}_{0}+\frac{1}{\zeta}p^{c}_{0}\left(1-e^{-\frac{\zeta}{M}t}\right).
\end{eqnarray}
We observe that the mean momentum decreases exponentially with time (with the decaying time scale $\tau_{B}=M/\zeta$) and the particle position is shifted by $p^{c}(0)/\zeta$ at long-times. 

We now consider for simplicity that $p^{c}_{0}=0$ and $y^{c}_{0}=0$. Other physical quantities of interest are the mean quadratic momentum, i.e., $\left<p^{c}(t)p_{c}(t)\right>$, and the mean square displacement (MSD), $s^{2}=x^{c}x_{c}$. In order to calculate both, it is useful to find the temporal correlation function between two momenta, $p^{a}(t)$ and $p^{b}(t^{\prime})$, at times $t$ and $t^{\prime}$, given by (see Appendix B for further details)
\begin{eqnarray}
\label{pp2}
\left<p^{a}(t)p^{b}(t^{\prime})\right>=\frac{M}{2\zeta}\Omega\delta^{ab}\left[e^{-\frac{\zeta}{M}\left|t-t^{\prime}\right|}-e^{-\frac{\zeta}{M}\left(t+t^{\prime}\right)}\right].
\end{eqnarray} 
Using previous equation, it is straightforward to obtain the mean quadratic momentum:
\begin{eqnarray}
\label{momentocuadrado}
\left<p^{c}(t)p_{c}(t)\right>=\frac{d\Omega M}{2\zeta}\left(1-e^{-2\frac{\zeta}{M}t}\right).
\end{eqnarray}
Proceeding along the same lines, one can straightforwardly derive the MSD:
\begin{eqnarray}
\left<s^{2}(t)\right>=\frac{d\Omega M}{\zeta^{3}}\left[\frac{\zeta}{M}t-\frac{1}{2}(e^{-2\frac{\zeta}{M}t}-1) \right. \nonumber \\ \left. -2(1-e^{-\frac{\zeta}{M}t})\right].
\label{planesol}
\end{eqnarray}

In the diffusive regime, $t\gg \tau_{B}$, the average kinetic energy reaches its equilibrium value. This allows us the evaluation of $\Omega$ from the equipartition theorem. Thus, $\left<p^{c}(t)p_{c}(t)\right>=d k_{B}T/2$ and $\Omega=2\zeta k_{B}T$, where $k_{B}$ is the Boltzmann constant and $T$ the absolute temperature. We also observe that in this time regime the MSD reproduces the standard kinematical Einstein relation  $\left<s^{2}(t)\right>=2dD_0 t$, where $D_0 =k_{B}T/\zeta$ is the free-particle diffusion coefficient \cite{Dhont}. We should point out that the value of $\Omega$ is independent of whether the space is curved or not, since it only depends on quantities intrinsic to the fluid, as solvent friction and particle dimension.

Higher order temporal correlation functions are also useful. In particular, we will see below that the four-point function $G^{abcd}(t_{1},t_{2},t_{3},t_{4})\equiv \left<p^{a}(t_{1})p^{b}(t_{2})p^{c}(t_{3})p^{d}(t_{4})\right>$ is necessary in order to obtain the leading curvature corrections. This correlation function can be computed by using the Wick's theorem \cite{Zinn-Justin},
\begin{eqnarray}
\label{G4}
G^{abcd}(t_{1},t_{2},t_{3},t_{4})&=&\left<p^{a}(t_{1})p^{b}(t_{2})\right>\left<p^{c}(t_{3})p^{d}(t_{4})\right>\nonumber\\&+&\left<p^{a}(t_{1})p^{c}(t_{3})\right>\left<p^{b}(t_{3})p^{d}(t_{4})\right>\nonumber\\&+&\left<p^{a}(t_{1})p^{d}(t_{4})\right>\left<p^{b}(t_{2})p^{c}(t_{3})\right>.
\end{eqnarray} 

\subsubsection{Leading weak curvature effects}

We turn now to the derivation of the leading weak curvature effects on the particles dynamics. As we already discussed, the Langevin equation is quadratic in the momentum and that contribution is coupled to the particles positions through the Chrystoffel symbols. The resulting equations are difficult to solve analytically, among other reasons because the left-hand side of equation (\ref{LocalL1}) involves a temporal derivative of the metric. Using $p^{a}=g^{ab}p_{b}$ ($p_{a}$ is independent of the metric), the local Langevin equation allows us to obtain the following expressions,
\begin{eqnarray}
\dot{p_{d}}&=&-\frac{\zeta}{M}p_{d}-\frac{1}{M}g_{cd}\left(\partial_{a}g^{cf}\right)g^{ab}p_{b}p_{f} \nonumber \\
&&-\frac{1}{M}g_{cd}\Gamma^{c}_{~ba}g^{bf}g^{ah}p_{f}p_{h}+f_{d},\nonumber\\
\dot{x^{a}}&=&\frac{1}{M}g^{ab}p_{b}.
\label{Sing}
\end{eqnarray}

In order to explore curvature effects, we expand equation (\ref{Sing}) around the planar solution (\ref{pplanar}). To reach this goal, we use the Riemann normal coordinates \cite{muller}. In normal coordinates, we have
\begin{eqnarray}
\label{eqRT}
g^{ab}&=&\delta^{ab}-\frac{1}{3}R^{a}_{~cd}{}^{b}x^{c}x^{d}+O(x^{3}),\nonumber\\
\Gamma^{c}_{~ba}&=&\frac{1}{3}\left(R^{c}_{~bda}+R^{c}_{~adb}\right)x^{d}+O(x^{3}),
\end{eqnarray}
where $R^{a}_{~bcd}$ are the components of the Riemann curvature tensor. Using (\ref{eqRT}) in (\ref{Sing}) one obtains
\begin{eqnarray}
\dot{p_{d}}&=&-\frac{\zeta}{M}p_{d}-\frac{1}{3M}R_{dbfa}x^{f}p^{b}p^{a}+\cdots+f_{d}\nonumber\\
\dot{x^{a}}&=&\frac{1}{M}(\delta^{ab}-\frac{1}{3}R^{a}_{~cd}{}^{b}x^{c}x^{d}+\cdots)p_{b}.
\label{ApproxRC}
\end{eqnarray}
We should notice that the Langevin equation in Euclidean geometries (\ref{euclideang}) is recovered when the curvature vanishes. In order to find a solution around the Euclidean case (\ref{pplanar}), we expand the momentum and position in the following way: $p_{d}=q_{d}+\delta{q}_{d}$ and $x^{a}=z^{a}+\delta{z}^{a}$, where $q_{d}$ and $z^{a}$ are the solutions for zero curvature, given by (\ref{pplanar}). Here, we have assumed that $\delta q =0$ and $\delta z=0$ when $R^{a}_{~bcd}=0$. If we consider only linear terms in curvature, we obtain the equation for $\delta{q}_{d}$,
\begin{eqnarray}
\dot{{\delta q}_{d}}=-\frac{\zeta}{M}{\delta q}_{d}-\frac{1}{3M}R_{dbfa}z^{f}q^{b}q^{a}.
\label{postime}
\end{eqnarray}
The second term of the right-hand side does not depend on $\delta{q}$; it depends only on time. The integration of equation (\ref{postime}) is similar to the one in the planar case. The initial condition for $\delta{q}_{d}(t)$ is $\delta{q}_{d}(0)=0$, since $p_{d}(t)$ satisfies $p_{d}(0)=q_{d}(0)$. Therefore, the momentum, up to linear terms, in an arbitrary Riemannian geometry is given by
\begin{eqnarray}
p_{d}(t)&=&q_{d}(t)-\frac{1}{3M^{2}}R_{dbca}\int_{0}^{t^{\prime}}dt^{\prime}e^{-\frac{\zeta}{M}(t-t^{\prime})}\nonumber\\
&&\times\int_{0}^{t^{\prime\prime}}dt^{\prime\prime}q^{c}(t^{\prime\prime})q^{b}(t^\prime)q^{a}(t^{\prime}),
\end{eqnarray}
and the position, up to linear terms as well, takes the form
\begin{eqnarray}
x^{a}(t)&=&z^{a}(t)-\frac{1}{3M^{3}}R^{a}_{bcd}\int_{0}^{t}dt^{\prime}\int_{0}^{t^{\prime\prime}}dt^{\prime\prime}e^{-\frac{\zeta}{M}(t^{\prime}-t^{\prime\prime})}\nonumber \\ 
&&\times \int_{0}^{t^{\prime\prime}}dt^{\prime\prime\prime}q^{b}(t^{\prime\prime})q^{c}(t^{\prime\prime\prime})q^{d}(t^{\prime\prime}).
\end{eqnarray}
Wick's theorem allows us to determine the temporal correlation functions of $q_{a}(t)$. Therefore, we have found that the odd correlations vanish, as in the case of the mean values of the momentum and position: $\left<p_{a}(t)\right>=0$ and $\left<x_{a}(t)\right>=0$. This means that there is not preferential points on the surface and the mean values are independent of the geometry. This result may change however for non-zero initial conditions.

Up to linear terms in the curvature, we obtain the following expectation value for $s^{2}(t)$
\begin{eqnarray}
\label{MSD}
\left<s^{2}(t)\right>=\left<z^{a}(t)z_{a}(t)\right>-\frac{1}{3M^{4}}R^{a}_{~\left(ijk\right)}J_{a}^{~ijk}(t)+\cdots,
\end{eqnarray}
where $\left<z^{a}(t)z_{a}(t)\right>$ is the same as in equation (\ref{planesol}), $R^{a}_{~\left(ijk\right)}\equiv R^{a}_{~ijk}+R^{a}_{~kji}$ and
\begin{eqnarray}
J_{aijk}(t)&=&\int_{0}^{t}dt_{1}\int_{0}^{t}dt_{2}\int_{0}^{t_{2}}dt_{3}\int_{0}^{t_{3}}dt_{4}e^{-\frac{\zeta}{M}(t_{2}-t_{3})}\nonumber\\
&&\times G_{aijk}(t_{1},t_{3},t_{4},t_{3}).
\end{eqnarray}
The quantity $J_{aijk}(t)$ captures the dynamical contribution that appears in the weak curvature regime. In addition, the four-point correlation function $G_{aijk}$ is defined according to equation (\ref{G4}); it is built by the products of two-point correlation functions and each of them carries a Kronecker's delta. Hence, using the symmetries of the Riemann tensor, the MSD reduces to
\begin{eqnarray}
\label{MSD}
\left<s^{2}(t)\right>=\left<z^{a}(t)z_{a}(t)\right>-\frac{2R_{g}}{3}J(t)+\cdots,
\label{result}
\end{eqnarray}
where $R_{g}$ is the Ricci scalar curvature and
\begin{eqnarray}
\label{Jexp}
J(t)&=&\frac{1}{M^{4}}\int_{0}^{t}dt_{1}\int_{0}^{t}dt_{2}\int_{0}^{t_{2}}dt_{3}\int_{0}^{t_{3}}dt_{4}e^{-\frac{\zeta}{M}(t_{2}-t_{3})}\nonumber\\
&&\times\left(G(t_{1},t_{4})G(t_{3},t_{3})-G(t_{1},t_{3})G(t_{4},t_{3})\right).
\label{Jfunction}
\end{eqnarray}
Equation (\ref{Jexp}) can be straightforwardly integrated (see Appendix B). Equation (\ref{result}) represents the MSD (geodesic mean square displacement) in the weak curvature regime. As we can appreciate from equation (\ref{Jexp}), the time scale $\tau_{B}=M/\zeta$ defines two time regimes: The one with $t\ll \tau_{B}$ (but very much larger than $\tau_{solvent}$) or the ballistic regime, and the one with $t\gg \tau_{B}$ called the diffusive regime. In the first case, the MSD is given by
\begin{eqnarray}
\left<s^{2}\left(t\right)\right>\approx 2d\frac{k_{B}T\zeta}{M^{2}}t^{3}-\frac{52}{15}\left(\frac{k_{B}T\zeta}{M^2}\right)^2 R_{g} t^{6}
\end{eqnarray}
The cubic term is the ordinary contribution to the ballistic regime when the initial condition is $p^{c}_{0}=0$ (it becomes quadratic in $t$ for non-zero initial conditions \cite{Dhont}). The next curvature contribution is of order $t^{6}$; typically negligible unless there is a region of very high curvature. 

In the  diffusive regime, $t\gg \tau_{B}$,  the function $J\left(t\right)$ reduces to $J\left(t\right)\approx \left(D_0 t\right)^{2}$. Therefore, the MSD becomes
\begin{eqnarray}
\label{MSDCurv}
\left<s^{2}(t)\right>=2dD_0 t-\frac{2R_{g}}{3}\left(D_0 t\right)^{2}+\cdots.
\end{eqnarray}
This result is the same found by one of us \cite{Castro} by means of the diffusion equation on curved manifolds. The MSD shows a deviation from the planar result due to curvature effects. Furthermore, equation (\ref{MSDCurv}) also shows the raise of two different diffusive regimes: The one with $\tau_{B} \ll t < \tau_{G}$, and the overdamped regime, also called geometric regime, $t \gg \tau_{G}$. Here, $\tau_{G}=3d/\left|R_{g}\right|D_0$ stands for the time thereafter the curvature effects become dominant and it is the regime when the equilibrium is reached. This result is a confirmation that the Langevin equation describes the same dynamics of the diffusion equation on curved manifolds in the diffusive regime. It is noteworthy to mention that this result has been recently obtained, using alternative methods, by M. Polettini \cite{Polettini}.

It is important to mention that in the planar case, i.e., $\left|R_{g}\right|\rightarrow 0$, the particle cannot feel any effect associated with the geometry ($\tau_{G}$ is never reached, then it grows towards infinity). Additionally, we should emphasize that in the particular case of $d=1$ the MSD may exhibit deviations from the planar result that cannot be associated to $R_{g}$, since the Gaussian curvature of lines is zero. In fact, as we will see further below, those effects are associated with the finite-size of the phase space.

From now on, we use the fact that Langevin equation and diffusion equation on curved manifolds describe the same dynamics in the diffusive and geometric regime. In the following section, we explicitly discuss some properties of the diffusive motion of the particles along a circle, $S^{1}$, and on a sphere, $S^{2}$.

\section{Diffusion in $S^{1}$ and $S^{2}$}

We now choose the diffusion equation in order to study the geometric regime ($t\gg\tau_{G}$) in the manifolds $S^{1}$ and $S^{2}$ (for a discussion on the diffusion on arbitrary hyperspheres see, for example, Ref. \cite{Jean}). The diffusion equation on curved manifolds can be written as,
\begin{eqnarray}
\frac{\partial P\left(x,x^{\prime},t\right)}{\partial t}&=&D_0 \Delta_{g}P\left(x,x^{\prime},t\right),\nonumber\\
P\left(x,x^{\prime},0\right)&=&\frac{1}{\sqrt{g}}\delta^{\left(d\right)}\left(x-x^{\prime}\right),
\label{difeq}
\end{eqnarray}
where $P\left(x,x^{\prime},t\right)dv$ is the probability of finding the diffusing particles in the volume element $dv=\sqrt{g}d^{d}x$, given that they began to move at $x^{\prime}$. The probability density distribution $P\left(x,x^{\prime},t\right)$ is normalized with respect to the volume $v$ of the manifold and $D_0$ is the free-particle diffusion coefficient. The operator $\Delta_{g}$, called the Laplace-Beltrami operator, is defined by
\begin{eqnarray}
\Delta_{g}f=\frac{1}{\sqrt{g}}\partial_{a}\left(\sqrt{g}g^{ab}\partial_{b}f\right),
\label{L-Bop}
\end{eqnarray}
with $g=\det\left( g_{ab}\right)$ and $f$ is a scalar function. The geometry is coupled to the Brownian motion through the me\-tric. It is clear that $P\left(x,x^{\prime},t\right)$ reaches a constant value when the system is under equilibrium conditions, i.e., $t\rightarrow\infty$. The diffusion equation (\ref{difeq}) is the same as the heat kernel equation and it has a lot of applications in the context of field theories on curved spaces \cite{Vassilevich}. 

The expectation value of a scalar function $\mathcal{O}$ defined on the manifold is given in the standard fashion, i.e.,
\begin{eqnarray}
\left<\mathcal{O}\left(x\right)\right>=\int_{\mathbb{M}} dv~\mathcal{O}\left(x\right)P\left(x,x^{\prime},t\right),
\end{eqnarray}
and $\left<\mathcal{O}\left(x\right)\right>$ depends on the initial point $x^{\prime}$. The characteristics of observables in manifolds are related with the particular structure of $P\left(x,x^{\prime},t\right)$. Besides, the probability density distribution $P\left(x,x^{\prime},t\right)$ can be determined by solving the eigenvalue problem $-\Delta_{g}\Psi=E\Psi$, where $E$ is the eigenvalue corresponding to the eigenfunction $\Psi$. In addition, it is known that for compact manifolds, the spectra of $\Delta_{g}$ is discrete and it can be written in a growing sequence $\left\{\lambda_{0}=0, \lambda_{1}, \lambda_{2}, \dots\right\}$, where $\lambda_{I+1}>\lambda_{I}$ \cite{chavel}. We also have a sequence of orthogonal eigenfunction $\Psi_{1}, \Psi_{2}, \cdots$ in $L^{2}\left(\mathbb{M}\right)$ (square-integrated functions of $\mathbb{M}$). In this sense, the probability density distribution can be formally  written as \cite{Grygorian} 
\begin{eqnarray}
P\left(x,x^{\prime},t\right)=\sum_{I}e^{-\lambda_{I}D_0 t}\Psi^{*}_{I}\left(x^{\prime}\right)\Psi_{I}\left(x\right),
\end{eqnarray}
with $\Psi^{*}$ being the complex conjugate of $\Psi$. We note that degeneracy of eigenvalues is explicitly considered in the sum. 

Now, let us consider an arbitrary observable $\mathcal{O}$. Its dynamical behavior can be obtained using the formal expression for $P\left(x,x^{\prime},t\right)$. The expectation value $\left<\mathcal{O}\left(x\right)\right>$ has a generic form; its structure around the geometric regime is determined by the smallest eigenvalues. Then, it can be written as follows:
\begin{eqnarray}
\label{obev}
\left<\mathcal{O}\left(x\right)\right>\approx \frac{1}{v}\int dv ~\mathcal{O} -a_{1}e^{-D_0 t \lambda_{1}}+\cdots,
\end{eqnarray}
where $a_{1}=\frac{1}{v}\int dv~\Psi^{*}_{1}\mathcal{O}\Psi_{1}$ (it is also convenient to define $a_{0}=\frac{1}{v}\int dv ~\mathcal{O}$).  It is remarkable that the leading term obtained in this way allows us to determine the steady spacial density
\begin{eqnarray}
P^{*}\left(x\right)=\frac{1}{v}\sqrt{g\left(x\right)}.
\end{eqnarray}
This is also consistent with the original calculation by Kramers \cite{Kramers} and recently discussed in \cite{Polettini}. We can easily obtain some properties of any observable by looking at the particular form of equation (\ref{obev}). For example, at long times the expectation value $\left<\mathcal{O}\left(x\right)\right>$ becomes $a_{0}$ as a consequence of the finite size of the space.  In physical terms, every observable that depends on  the position will remain fixed, on average, and its distribution does not longer evolve with time.  The quantity $a_{0}$ is the geometrical average of $\mathcal{O}$; this is the reason we called this regime the geometric regime. Although counterintuitive, the values of the observables do not depend on the temperature for $t\gg\tau_G$; it is only a function of the surface geometry. The value $a_{0}$ is also the mean-value in the equilibrium regime. This result is, indeed, the ge\-ne\-ra\-li\-za\-tion to curved space of a classical ideal gas in the three-dimensional Euclidean space $\mathbb{R}^{3}$. 

\subsection{Brownian motion over $S^{1}$}

Brownian motion on the circle represents, after the motion on the straight line, the simplest example where there is a clear manifestation of the geometrical effects on the particle dynamics, but it is also the most fundamental one, since it is fully described by a single physical variable. It is also relevant for the theoretical and experimental study of single-file diffusion in quasi-one-dimensional interacting systems (see, e.g., \cite{Chava2010} and references therein).

The circle is the mappping ${\bf X}:\left[0,2\pi\right]\to\mathbb{R}^{2}$, where ${\bf X}=\left(R\cos\varphi, R\sin\varphi\right)$, with $R$ being the circle radius. The Laplace-Beltrami operator in this case takes the form $\Delta_{S^{1}}=\frac{1}{R^{2}}\frac{\partial^{2}}{\partial\varphi^{2}}$. The eigenfunctions of this operator form the complete orthonormal set $\left\{e^{i m\varphi}\left.\right| m\in\mathbb{Z}\right\}$ in $L^{2}(S^{1})$ and their corresponding eigenvalues are $\lambda_{m}=-m^{2}/R^{2}$. 

In order to study Brownian motion on $S^{1}$, we choose the following initial and boundary conditions: $\varphi^{\prime}(0)=0$ and  $P\left(\varphi, 0, 0\right)=\delta\left(\varphi\right)/2\pi R$. After some simplifications, the explicit solution of the diffusion equation is
\begin{eqnarray}
P(\varphi,t)=\frac{1}{2\pi R}\left(1+2\sum_{m=1}^{\infty}e^{-m^{2}\frac{D_0 t}{R^{2}}}\cos\left(m\varphi\right)\right).
\end{eqnarray}
In this case, the distribution is normalized with the perimeter of the circle, i.e., $\int_{I}ds~ P(\varphi,t)=1$, where $I=(-\pi,\pi)$ and $ds=R~d\varphi$. The distribution is also symmetric under the interchange $\varphi\to-\varphi$.

The first moment,  $\left<s(t)\right>$, and the second moment or MSD, $\left< s^{2}(t)\right>$, of the distribution can be straightforwardly evaluated. The former is zero, since the distribution is an even function, whereas the MSD has the form
\begin{eqnarray}
\label{MSD}
\frac{\left< s^{2}(t)\right>}{R^{2}}=\frac{\pi^{2}}{3}+4\sum^{\infty}_{m=1}\left(-1\right)^{m}\frac{e^{-m^2 \frac{D_0 t}{R^2}}}{m^{2}},
\label{MSDcirculo}
\end{eqnarray}
with $s=R\varphi$ being the arc-length. On the one hand, the MSD given by equation (\ref{MSDcirculo}) reduces to $\left< s^{2}(t)\right>=2D_0 t$ for short times ($\tau_B \ll t < \tau_G$). On the other hand, at long times ($t\gg t_{G}=R^{2}/D_0 $) we have $\left< s^{2}(t)\right>=\pi^{2}R^{2}/3$. In the geometric regime the dependence is only on the size of the circle. The numerical evaluation of equation (\ref{MSDcirculo}) is shown in figure 1.

As we mentioned previously, although the MSD in (\ref{MSDcirculo}) deviates from the planar result, this difference is due to the finite size of the circle and not to curvature effects. We compare the predictions of equation (\ref{MSD}) with computer simulation results in figure 1. The latter ones will be explained further below.

\begin{figure}[t]
\begin{center}
\label{fig1}
\includegraphics[width=8cm]{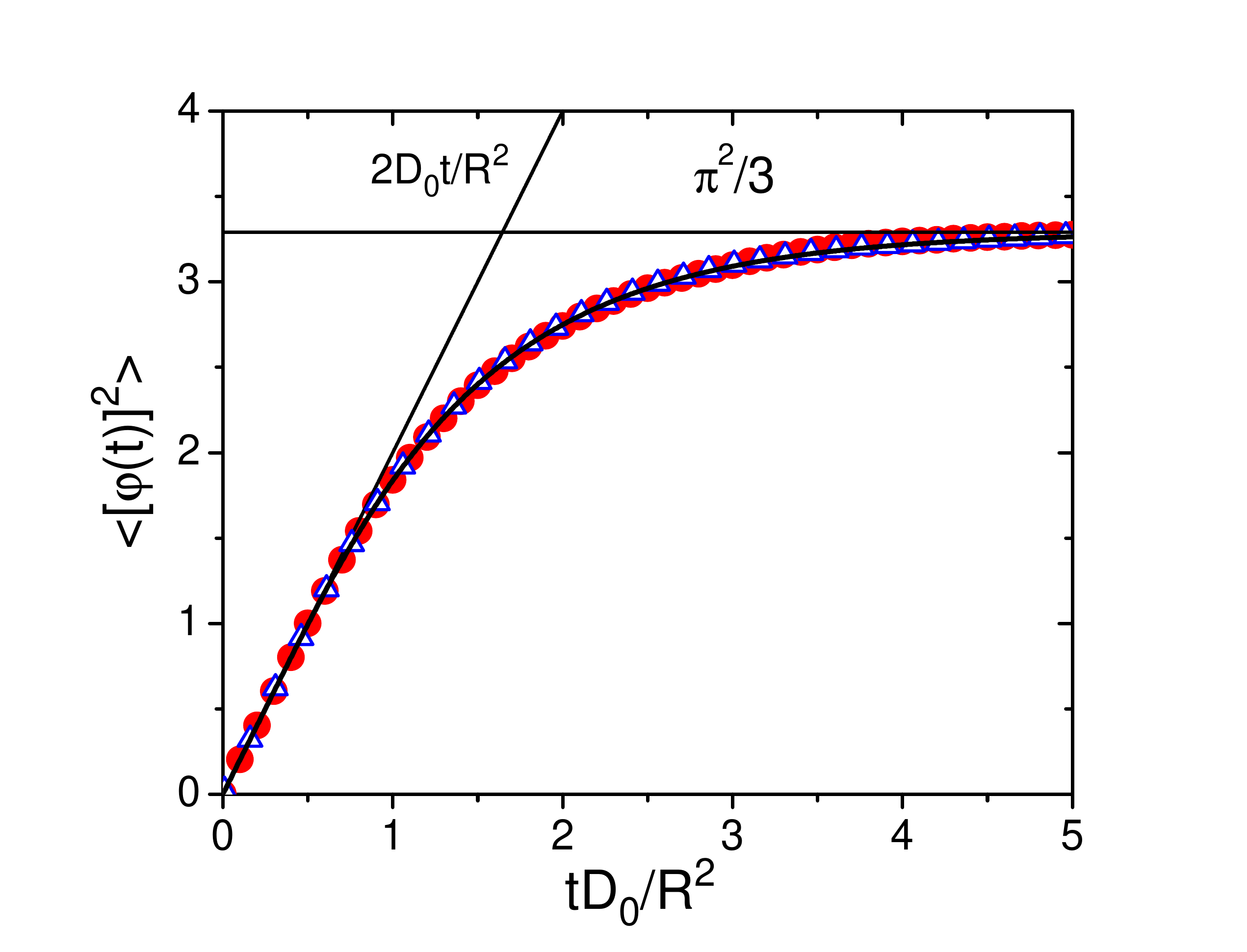}
\caption{Mean square angular displacement as a function of time for free Brownian particles diffusing along a circle. The line corresponds to our theoretical result given by equation (\ref{MSDcirculo}), and the symbols to the Brownian computer simulations results obtained by means of both the standard Ermak and McCammon algorithm (circles) and its heuristic adaptation to curves (triangles). The error bars of the simulation data are smaller than the size of the symbols. There is no appreciable difference between the results. The straight lines stand for the short and long-time limits as indicated.}
\end{center}
\end{figure}

\subsection{Brownian motion over $S^{2}$}

We now study the Brownian dynamics on the sphere putting special emphasis in the geometric regime. It is noteworthy to mention that several features of this special case have already been studied by several authors \cite{Holyst,faraudo02,K,Tomoyoshi,muthu98, ghosh} and it was originally used to study the rotational Brownian dynamics of rods within the Debye theory \cite{Dhont, DifRot}, where non-interacting rods can be cast into a diffusion equation on the unit sphere. Here, this special case is emphasized in the geometric regime where we use the geodesic distance as the displacement of the particle as in Ref. \cite{faraudo02}. In the sphere, the geodesic distance corresponds to a section of one Riemann great circle.  The geometry of a sphere is encoded into the metric given by
\begin{eqnarray}
ds^2\equiv g_{ab}dx^{a}dx^{b}=R^2\left(d\theta^2+\sin^2\theta d\varphi^2\right),
\end{eqnarray}
where $R$, $\theta$ and $\varphi$ are the radius, polar and azimuthal coordinates of the sphere, respectively. The Laplace-Beltrami operator on the sphere has eigenvalues and eigenvectors given by $\lambda_{\ell}=\ell\left(\ell+1\right)$ and $\left\{Y_{\ell m}\left(\theta,\varphi\right)\right\}$ with $\ell=0,\cdot\cdot\cdot, \infty$ and $m=-\ell,\cdot\cdot\cdot,\ell$; $Y_{\ell m}\left(\theta,\varphi\right)$ being the standard spherical harmonics.  

We choose $x^{\prime}$ to be on the north pole and take advantage of the rotational invariance. Besides, the boundary condition (\ref{difeq}) is explicitly taken into account. The solution of the diffusion equation is then
\begin{eqnarray}
P\left(\theta,t\right)=\sum^{\infty}_{\ell=0}\frac{2\ell+1}{4\pi R^{2}}~P_{\ell}\left(\cos\theta\right)\exp\left[-\frac{D_0 \ell\left(\ell+1\right)}{R^2}t\right],
\label{EqProb}
\end{eqnarray}
where $P_{\ell}$ is the Legendre polynomial of order $\ell$. As in the previous case, we look for the information provided by $\left<s(t)\right>$ and $\left<s^{2}(t)\right>$, but we have now that $s=R\theta$.

By means of the operator method defined in \cite{Castro}, it is possible to show that the short-time behavior of the MSD is given by equation (\ref{MSDCurv}) with the Gaussian curvature of the sphere, $R_{g}=2/R^{2}$. It is interesting to note that the terms in the MSD that depend on the Gaussian curvature are always negative. This means that curvature effects only contribute to reduce the particle diffusion with time.

In the geometric regime, $t\gg\tau_{G}=3R^{2}/D_0$, we obtain from equation (\ref{obev}) the following expressions,
\begin{eqnarray}
\label{limitcase}
\frac{\left<s\right>}{R}&=&\frac{\pi}{2}\left(1-\frac{3}{4}e^{-2\frac{D_0 t}{R^2}}+\cdot\cdot\cdot\right)\nonumber\\
\nonumber\\
\frac{\left<s^2\right>}{R^2}&=&\frac{\pi^2-4}{2}\left(1-\frac{3\pi^2}{4\pi^2-16}e^{-2\frac{D_0 t}{R^2}}+\cdot\cdot\cdot\right).\nonumber\\
\end{eqnarray}
At the beginning the particles move around their initial position, i.e., the north pole. After a long time, very much larger than $\tau_G$, the expectation values $\left<s\right>$ and $\left<s^2\right>$ move towards the saturation values $\pi R/2$ and $(\pi^2-4)R^2/2$, respectively. The particle has visited all the points on the surface and confinement dominates entirely the diffusive behavior; the saturation values only depend on the size of the sphere. The behavior of equation (\ref{limitcase}) is shown in figure 2.

\begin{figure}[t]
\begin{center}
\label{fig2}
\includegraphics[width=8cm]{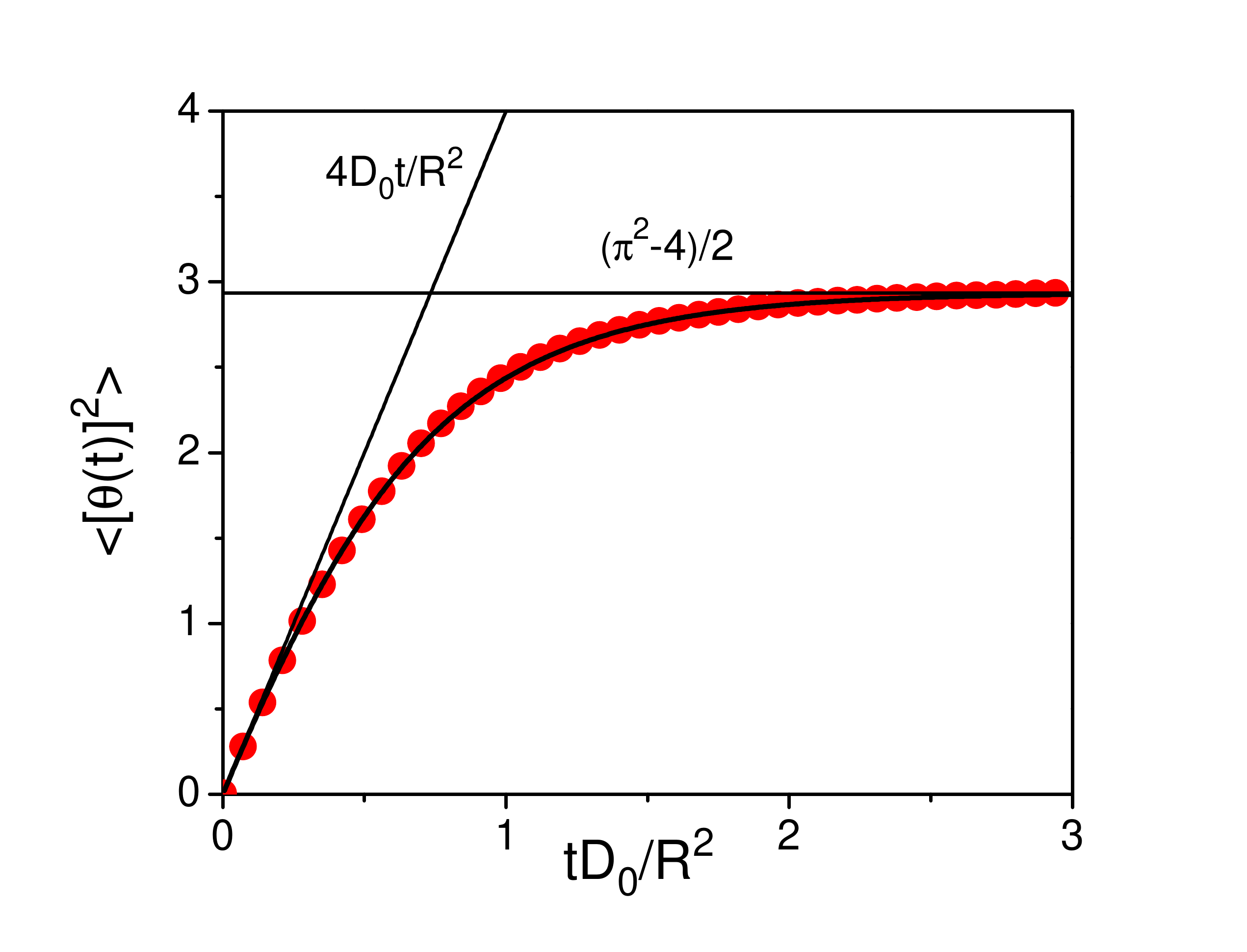}
\caption{Mean square polar angular displacement as a function of time for free Brownian particles diffusing over a sphere. The line correspond to our theoretical result given by equation (\ref{limitcase}), and the symbols to the Brownian computer simulations results obtained by means of the standard Ermak and McCammon algorithm (circles). The error bars of the simulation data are smaller than the size of the symbols. There is no appreciable difference between the results. The straight lines stand for the short and long-time limits as indicated.}
\end{center}
\end{figure}

The expectation value of any observable $\mathcal{O}=\mathcal{O}(\theta,\varphi)$ on the sphere can be written as
\begin{eqnarray}
\label{eqomega}
\left<\mathcal{O}\left(\theta,\varphi\right)\right>=\frac{R}{2}\sum^{\infty}_{\ell=0}\left(2\ell+1\right)g_{\mathcal{O}}\left(\ell\right)e^{-D_0 \ell\left(\ell+1\right)t/R^2}, 
\label{expectation}
\end{eqnarray}
where $g_{\mathcal{O}}$ is the projection of $\mathcal{O}$ along the basis of Legendre polynomial. We explicitly show the functional form of $g_{\mathcal{O}}$ in Appendix C, for both $g_{s}$ and $g_{s^{2}}$.

\section{Brownian dynamics simulations on curved surfaces} 

\subsection{Standard Ermak and McCammon algorithm}

In 1978, Ermak and McCammon introduced a method for simulating the Brownian dynamics of particles \cite{Ermak}. This method, which has been adapted in Euclidean coordinates, was derived from the Langevin equation and became consistent with the Fokker-Planck equation. Furthermore, such a method can be straightforwardly applied when either hydrodynamic interactions are con\-si\-dered explicitly or external forces act on the particles. This method has been successfully employed to study the structural and dynamic properties of a large variety of complex fluids, i.e., colloids, polymers, etc. \cite{Duenweg}

The algorithm of Ermak and McCammon \cite{Ermak} is given by
\begin{equation}
\mathbf{X}_{\alpha}=\mathbf{X}_{\alpha}^{0} + \sum\limits_{\beta=1}^{N} \frac{\partial \mathbf{\mathbf{D}}_{\alpha \beta}^{0}}{\partial r_{\beta}} \Delta t + \sum\limits_{\beta=1}^{N} \beta \mathbf{\mathbf{D}}_{\alpha \beta}^{0} \mathbf{F}_{\beta}^{0}\Delta t + \delta \mathbf{X}_{\alpha},
\label{eermak}
\end{equation}
where $N$ is the number of particles, $\beta = (k_{B} T)^{-1}$ is the inverse of the thermal energy. The hydrodynamic interactions (HI) are included through the diffusion tensor $\mathbf{D}_{\alpha \beta}^{0}$, $\mathbf{F}_{\beta}^{0}$ is the total force exerted on the $\beta$-th particle and the index $0$ tells us that the variable must be calculated at the beginning in time at every step. The term $\delta \mathbf{X}_{\alpha}$ represents a random displacement with a Gaussian distribution function with mean value zero and a covariance matrix given by the elements $\left\langle \delta X_{\alpha_i} \delta X_{\beta_j}\right\rangle=2 D_{\alpha_i \beta_j}^{0} \Delta t$; these are the requirements needed to satisfy the fluctuation-dissipation theorem (\ref{fluc-dissip}). The indices $\alpha$ and $\beta$ run over the particles, and the indices $i$ and $j$ over the cartesian coordinates. In our case, we do not consider HI and, therefore, $D_{\alpha_i \beta_j}^0=\delta_{ij} \delta_{\alpha \beta}D_0$, where $D_0$ is again the free-particle diffusion coefficient. With this assumption, the second term in the right-hand side of equation (\ref{eermak}) disappears and allows us to simplify drastically the calculation of the third and fourth terms of the same side.

As we mentioned previously, the algorithm of Ermak and McCammon describes the temporal evolution of Euclidean variables. However, it can be still used to describe the dynamics of particles on curved surfaces. This can be done by considering an external field that constrains the movement of the particles on the surface. We demand that the force coming from such a field does not contribute to the tangent displacements of the particles, i.e., this force has to act normal to any point of the desired manifold (i.e. $S^{1}$ or $S^{2}$) at any time to guarantee that it does not perform work on the system. Then, the simplest vector force-field that satisfies such requirements can be written as
\begin{equation}
\mathbf{F}_{\alpha}=-\kappa(|{\bf X}_{\alpha}|-R)\mathbf{n}_{\alpha},
\label{forcer}
\end{equation}
where $\kappa$ is a coupling constant, whose value is chosen in such a way that the particle displacements in the perpendicular direction to the surface is basically negligible, $R$ is a parameter of this force that we identify, here,  with the radius of either the circle or the sphere and $\mathbf{{n}}_{\alpha}={\bf X}_{\alpha}/\left|{\bf X}_{\alpha}\right|$ is a unit normal vector. This force can be thought as a spring-like force that attach the particle to a domain near the surface; in this sense $\kappa$ is a spring-like constant. In the two-dimensional case, this vector field can be explicitly visualized in figure (3), where the circle (solid line) shows the separation of the plane in two regions defined by the sign of $\mathbf{F}_{\alpha}$.
\begin{figure}[h]
\begin{center}
\label{fig41}
\includegraphics[width=5.5cm]{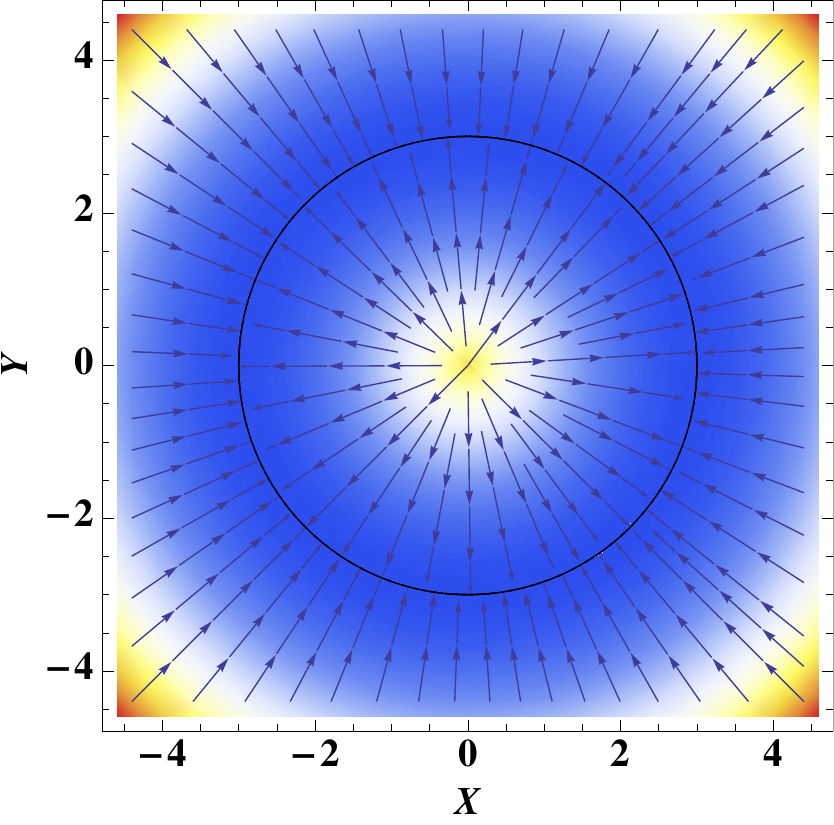}\\
\caption{For the two-dimensional case, the manifold $S^{1}$ separates the plane in two regions depending on the sign of the vector force-field ${\bf F}_{\alpha}$.}
\end{center}
\end{figure}
It is also convenient to determine the potential energy associate to this force. This is given by
\begin{eqnarray}
\label{potential}
V\left({\bf X}_{\alpha}\right)=\frac{k}{2}\left(\left|{\bf X}_{\alpha}\right|-R\right)^{2}.
\label{potential}
\end{eqnarray}
This potential has a ``Mexican hat" shape. In figure (4), we plot the potential given by equation (\ref{potential}) for the 2-dimensional case, where the points that minimize the potential correspond to the manifold (in this case $S^{1}$).
\begin{figure}[h]
\begin{center}
\label{fig31}
\includegraphics[width=6cm]{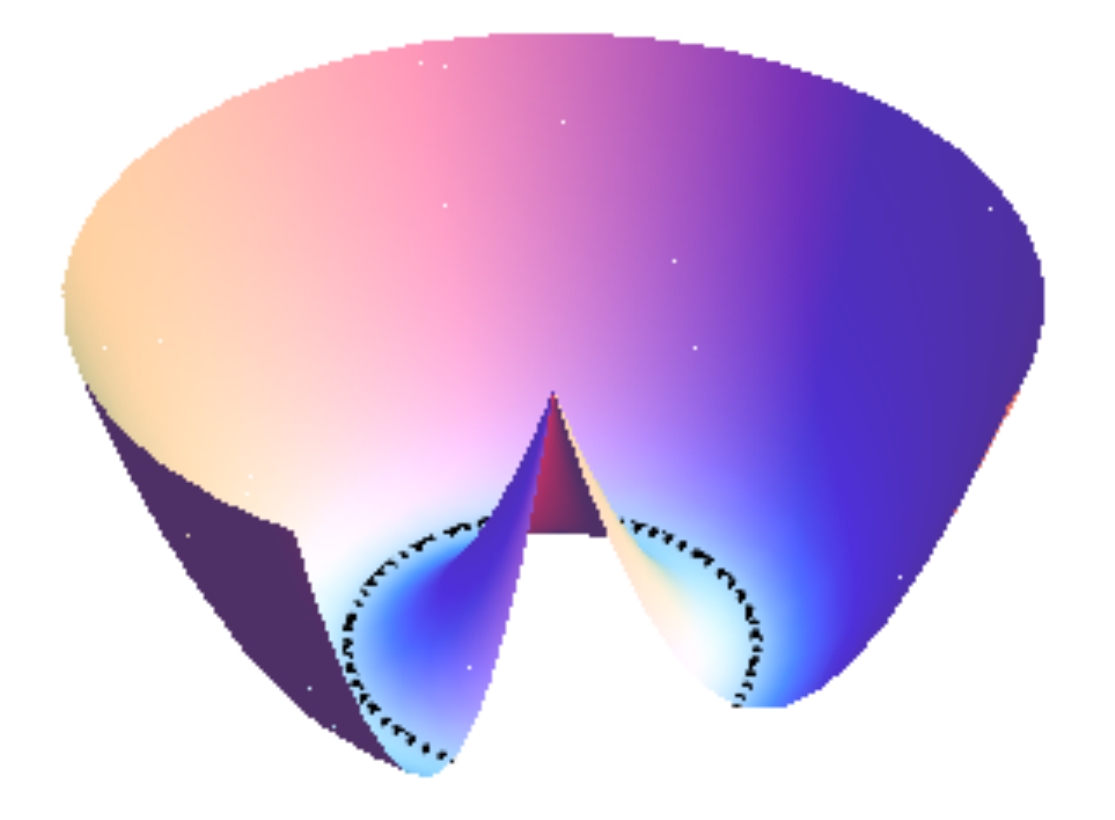}
\caption{For the two-dimensional case, the external potential (\ref{potential}) has a ``Mexican hat" shape, whose minima are the points of the manifold $S^{1}$. The value of the coupling parameter, $\kappa$, controls the width of the well.}
\end{center}
\end{figure}

Before we implement the external force (\ref{forcer}) in the Ermack-McCammon algorithm, let us provide two arguments that will help us to understand why it will reproduce the correct dynamics on either the circle or the sphere in the limit of very stiff potential, i.e. $\kappa\to\infty$. We analyze this limiting process by choosing the situation within the context of the Classical Mechanics and Statistical Mechanics. Thus, in the former case, we have a classical system consisting of a particle subjected to the vector force-field given by (\ref{forcer}). Following N. G. van Kampen and J. J. Lodder \cite{KampenandLodder} the motion of the particle will not have rapid vibrations in the normal direction of the surface, as far as the force acting upon the particle, that initiates its movement, vary smoothly during a short-time $\delta$ and when the condition $\delta\gg 1/\kappa^{1/2}$ is entirely satisfied. Under this assumption the result, as Kampen and Lodder pointed it out, is that the particle motion will be along the surface (in this case either $S^{1}$ or $S^{2}$).

Now, by using a Statistical Mechanics analysis, we perform the calculation of the single canonical partition function of the particle subjected to the external field and we compare it with the corresponding  partition function on the sphere. On one hand, the latter partition function is given by
\begin{eqnarray}
\mathbb{Z}_{S^{2}}\left(T\right)=\frac{A}{\lambda^{2}\left(T\right)},
\end{eqnarray}
where $A=4\pi R^{2}$ and $\lambda(T)=2\pi \hbar/\sqrt{2\pi M k_{B}T}$ is the de Broglie wavelength (see the appendix D for a derivation of this equation). On the other hand, the partition function for the particle subjected to the potential (\ref{potential}), after we integrate out the momenta, is given by
\begin{eqnarray}
\mathbb{Z}_{V}\left(k\right)=\frac{\mathcal{N}}{\lambda^{3}\left(T\right)}\int d^{3}{\bf x }\exp\left\{-\frac{\beta \kappa}{2}\left(\left|{\bf x}\right|-R\right)^{2}\right\}.
\end{eqnarray}
where $\mathcal{N}$ is some adimensional constant that does not change the physics. Naively, it is expected that for large values of $\kappa$ the only admissible value for ${\bf x}$ should be any position ${\bf x}\in \mathbb{R}^{3}$ with length $\left|{\bf x}\right|=R$ leading to the effect of confinement on the sphere.  However, this intuition is approximately correct because when temperature increase, the confinement effect would disappear. Thus, the only way to maintain this confinement is scaling the value of the coupling constant $\kappa$ with temperature in such a way that the particles are maintained on the sphere. Indeed, this happens as we show in the following. Using polar coordinates and performing several change of variables, last integral can be written as
\begin{eqnarray}
\label{Integral}
\mathbb{Z}_{V}=\frac{4\pi\mathcal{N}}{\lambda^{3}\left(T\right)}\left(\frac{2}{\beta\kappa}\right)^{\frac{3}{2}}\int^{\infty}_{-x_{0}}dy\left(y+x_{0}\right)^2\exp\left(-y^{2}\right),
\end{eqnarray}
where $x_{0}=\sqrt{\frac{\beta\kappa}{2}} R$. Last integral can be perfomed exactly in terms of the Error function (see appendix D) and for large value of $\kappa$ it has an asymptotic value that goes to $\sqrt{\pi} x^2_{0}$, therefore the partition function for $\kappa\to\infty$ is given by
\begin{eqnarray}
\mathbb{Z}_{V}\approx\frac{4\pi R^2}{\lambda^{3}\left(T\right)}\mathcal{N}\sqrt{\frac{2\pi}{\beta\kappa}}.
\end{eqnarray}
This means that the only way that this partition function converges to the one on the sphere ($\mathbb{Z}_{S^{2}}$) is such that the coupling $\kappa$ grows with temperature as
\begin{eqnarray}
\kappa\left(T\right)=\frac{k^2_{B}M\mathcal{N}^{2}}{\hbar^{2}}T^{2}.
\end{eqnarray}

The limit of large value of $\kappa$ can be performed for a fixed temperature taking the adimensional constant $\mathcal{N}\to\infty$. Therefore, taking this scaling law for the coupling constant $\kappa$, we have that $\mathbb{Z}_{S^{2}}=\lim_{\mathcal{N}\to\infty}\mathbb{Z}_{V}$. Thus, with these arguments in both Classical Mechanics and Statistical Mechanics we are confident that at least in these two extremal situations we can control the confinement effect of the particles on the surface by means of the stiffness parameter. 

Hence, equation (\ref{forcer}) is incorporated in the standard algorithm for Brownian dynamics described in equation (\ref{eermak}) to analyze the diffusion on the given surface. We should mention that the addition of force (\ref{forcer}) into equation (\ref{eermak}) has the same effect on the particle dynamics as the second term of the left-hand side in equation (\ref{LangP}), i.e., it only constrains the motion of the particles on the manifold. Thus, this kind of trick allows us to study the diffusion on curved surfaces (at least on $S^{1}$ and $S^{2}$) through the use of the standard Ermak and McCammon algorithm. It is noteworthy to mention that according to E. J. Hinch \cite{Hinch} in order to convert the Brownian motion of a rigid system to an equivalent very stiff system we have to add a {\it pseudo-corrective} force, ${\bf F}^{c}$, to the equation of motion. In the case of our interest, if we want to convert the Brownian motion of a very stiff system to an equivalent rigid system one would have to apply a force opposite to ${\bf F}^{c}$.  In the particular case of the sphere (similar for the circle) this force is given by ${\bf F}^{c}=-\frac{k_{B}T}{R}{\bf n}_{\alpha}$ (see appendix D for details). The correction term would be $-\kappa\left(\left|{\bf X}\right|_{\alpha}-R-\frac{k_{B}T}{\kappa R}\right){\bf n}_{\alpha}$ which does not have any contribution in the particular cases of sphere and circle.  Similar result, is indeed, already found by Grassia, Hinch, and Nitshe for the Brownian motion on an ellipse in \cite{Grassia}.

In our Brownian dynamics simulations, we have used $N=1000$ free particles, a reduced time step $\Delta t^{*}\equiv\Delta t D_{0}/R^{2}=10^{-5}$ and a reduced stiffness parameter $\kappa^{*}\equiv\beta\kappa R^{2}=10^{5}$. We also use $2\times10^{6}$ time steps to reduce the statistical uncertainties. Thus, the MSD from the adaptation of the standard algorithm of Ermak and McCammon is shown in figures 1 and 2 for $S^{1}$ and $S^{2}$, respectively.

\subsection{Heuristic adaptation of the Ermak and McCammon algorithm to curves}

Equation (\ref{eermak}) takes the simple form $\mathbf{X}_{\alpha}=\mathbf{X}_{\alpha}^{0} + \delta \mathbf{X}_{\alpha}$ for free particles, with $\left\langle \delta X_{\alpha_i} \delta X_{\alpha_j}\right\rangle=2  \delta_{i j}D_0 \Delta t$. In a $d$-dimensional Euclidian open space this process is equivalent to allow the particles to move in any direction with equal probability, as long as the distances they travel are Gaussian randomly distributed with variance $\left\langle \delta s_\alpha \delta s_\alpha \right\rangle=2 d D_0 \Delta t$. This is however the short-time behavior of the MSD in $d$-dimensional manifolds (\ref{MSDCurv}). Hence, we heuristically extend the Ermak and McCammon algorithm to curved manifolds by allowing the particles to move in any direction with equal probability, but the geodesic distances they travel are Gaussian randomly distributed, i.e., $s=s^{0} + \delta s$ with $\left\langle \delta s \delta s \right\rangle=2 d D_0 \Delta t$, as long as $\tau_{B} \ll \Delta t \ll \tau_{G}$.

In the particular case of a circle, this idea leads to the following algorithm: A uniform random number is generated in the interval $[0,1]$; the particle in turn is allowed to move in the clock-wise direction if the result falls in $[0,0.5]$, otherwise the particle moves in the opposite direction; a Gaussian randomly distributed number with variance $\left\langle \delta s \delta s \right\rangle=2 D_0 \Delta t$ is then generated in order to determine the arc-length the particle travels; these steps are repeated for every particle, many times, in order to construct the dynamics of the system in its natural sequence. In our simulations, we let 1000 free particles to move in very short time steps, until they approximately cover a distance of 100 times the perimeter of the circle. The large number of particles allows to improve the numerical precision of our results.

We expect, on the one hand, the short-time behavior $\left<\Delta s^{2}(t)\right>=2D_0 t$, since this is included in the construction of the algorithm. On the other hand, for very long times ($t\gg\tau_G$) the particles has to distribute uniformly along the perimeter of the circle. Therefore, the geometric behavior of the MSD must be given by the simple average of the geodesic square displacement
\begin{equation}
\frac{\left<\Delta s^{2}(t\gg\tau_G)\right>}{R^{2}}=\frac{1}{2\pi}\int_0^{2\pi} (\varphi-\left<\varphi\right>)^2 d\varphi=\frac{\pi^{2}}{3},
\label{circle1}
\end{equation}
which agrees with equation (\ref{MSDcirculo}). These and the intermediate values of $\left<\Delta s^{2}(t)\right>$ are shown in figure 1.

The extension of these ideas to the general case of curved surfaces will be presented elsewhere.

\section{Concluding remarks and perspectives}

In this work the diffusion of free particles on curved surfaces is studied. After writing the Langevin equation and the fluctuation-dissipation theorem for curved surfaces, we solved the former in the Riemann normal coordinates for weak curvatures, i.e., up to linear terms in the Riemann curvature tensor. From this solution, the dynamics of the particles can be clearly separated in three regimes; the ballistic one, $\tau_{solvent}\ll t \ll \tau_B$, and two diffusive regimes; short times, $\tau_{B}\ll t < \tau_G$, and long times, or geometric regime, $t \gg \tau_G$. In the ballistic regime we find effects of the geometry up to order of $t^{6}$ typically negligible unless there is a region of high curvature.  We therefore conclude that, typically, the local dynamics occurs in the plane tangent to the surface. Nevertheless, in the long-time diffusive regime only the geometric effects take place. The free particle diffusion coefficient $D_0$ might be understood in terms of the short-time limit of the mean geodesic square displacement, $\left<\Delta s^{2}(\tau_{B}\ll t \ll \tau_G)\right>=2dD_0t$, in a similar way as in the case of interacting particles. The geometry then appears as an external force acting on the diffusing par\-ti\-cles, which can be recognized in the second term of the left side of equation (\ref{LangP}).

We should remark that in the short-time diffusive regime the Langevin equation was found to have the same solution as the diffusion equation on curved surfaces \cite{Castro}, as it is expected and consistent with a work of M. Polettini \cite{Polettini}. We therefore used the latter in order to study the whole diffusive dynamics of free particles along a circle, $S^1$, and over a sphere, $S^2$. We do not expect curvature effects in $S^1$ since its Gaussian curvature $R_g$ is zero. However, the MSD displays a geometric diffusive regime due only to confinement effects, since the particles are unable to move beyond the region where the circle is placed. In $S^2$ the confinement and curvature effects act together to define the geometrical regime. The difference between curvature and confinement effects is subtle and somehow counterintuitive. This will be carefully reported somewhere else. 

We also reported some results from Brownian dynamics computer simulations. We obtained them by implementing the standard Ermak-McCammon algorithm, as well as its heuristic adaptation to curves. In the first case, we assumed that the particles are subjected to an external field that constrains the movement of the particle to the surface. The coupling constant $\kappa$ can be thought as a spring-like constant that is adjusted to guarantee the particle dynamics very close to the surface. A particular test of this field was made on the ground of Statistical Mechanics by calculating the single canonical partition function of the particle in the field and compare it with the corresponding partition function on $S^{2}$. It is found a curious effect, that may be experimentally tested, that the only way to maintain the confinement effect to the spherical surface is scaling the value of the coupling constant $\kappa$ with temperature in a precise way.  In the second case, which was only applied to the circle, we allowed the particles to move in every direction along the curve, every time displacing geodesic lengths given by random Gaussian number with variance $\left\langle \delta s \delta s \right\rangle=2D_0 \Delta t$. The quantitative comparison of the theoretical results with the simulation data was shown in figures 1 and 2.

Our approach can be extended in various directions. We could study the case of interacting particles where interaction may produce colored distributions for the stochastic forces in the Langevin equation \cite{GLEMedina}, so that Wick's theorem, which is of central importance in our calculations, were not longer valid. Nevertheless, it could be longer applied as an approximation, in the sense that the $n$-time correlation functions may be decomposed in terms of two-time correlation functions. In addition, both implementations of the Ermak and McCammon algorithm may be further used for interacting particles, as well as for other physical circumstances. For instance, the rotational Brownian motion of molecules can be studied by a diffusion equation on a manifold. For the case of the lateral diffusion of a protein or lipid  we did not take into account the effects of the thermal fluctuations of the membrane and on top of that the finite size of these particles could involve local deformations on the membrane that can change the diffusion constant. 
Furthermore, it could be interesting from the theoretical viewpoint to investigate if there are other ``realistic" circumstances where Lagrange constraints represent idealization of a very stiff potential.


\appendix

\setcounter{section}{0}

\section{Fluctuation-dissipation theorem} 

\noindent The stochastic force is Gaussian distributed for each point on the surface $S$. In global coordinates this distribution is given by \cite{Zinn-Justin}
\begin{eqnarray}
d\mu=\prod_{i=1}^{3}\left[df_{i}\right]\exp\left\{-\frac{1}{2\Omega}\int^{t}_{0} d\tau ~\delta^{ij}f_{i}\left(\tau\right)f_{j}\left(\tau\right)\right\},
\label{Distribution}
\end{eqnarray}
where $\left[df_{i}\right]$ is an appropiate functional measure. This is equivalent to a Gaussian vector field theory in one dimension. The expectation values are defined by $\left<\mathcal{O}\right>=\int d\mu~\mathcal{O}/\int d\mu$. In particular,  the fluctuation-dissipation theorem (\ref{fluc-dissip}) can be verified using (\ref{Distribution}).

The force distribution (\ref{Distribution}) also determines the fluctuaction-dissipation theorem in local coordinates (\ref{locfluc-diss}). 
To show this, let us separate the force in tangent and normal components. Since, ${\bf e}_{a}$ and ${\bf n}$ are given for each point, thus ${\bf f}={\bf e}^{a}f_{a}+{\bf n}f_{n}$ is a biyective transformation between $\left\{f_{i}\right\}$, with $i=1,2,3$, and $\left\{f_{a}, f_{n}\right\}$, with $a=1,2$. Thus the measure $\prod_{i=1}^{3}\left[df_{i}\right]$  transforms to $\prod_{a=1}^{2}\left[d{f}_{a}\right]\left[df_{n}\right]J$, where $J={\bf n}\cdot\left({\bf e}_{1}\times{\bf e}_{2}\right)$ is the Jacobian. In addition, the argument of the Boltzmann weight can be splitted in these coordinates. Then, the measure $d\mu$ can be written as
\begin{eqnarray}
d\mu=\prod_{a=1}^{2}\left[d{f}_{a}\right]\left[df_{n}\right]J  \exp \left\{-\frac{1}{2\Omega}\int dt \left(g^{ab}{f}_{a}{f}_{b}+f^{2}_{n}\right)\right\}.\nonumber\\
\label{Distribution2}
\end{eqnarray}
\noindent Now, since the hypersurface is locally a plane  we can always choose ${\bf e}_{a}$ such that $g_{ab}=\delta_{ab}$. Therefore, the local fluctuation-dissipation relations (\ref{locfluc-diss}) can be straightforwardly obtained from (\ref{Distribution}). This technical detail allows us to establish that both global and local versions of the Langevin equation on curved surfaces are equivalent.\\

\section{Correlation functions} 

\subsection{Green function}
\noindent The correlation of two momenta for zero initial conditions can be computed from 
\begin{eqnarray}
\left<p^{a}\left(t\right)p^{b}\left(t^{\prime}\right)\right>&=&e^{-\frac{t+t^{\prime}}{\tau_{B}}}\int^{t}_{0}dt_{1}\int^{t^{\prime}}_{0}dt_{2}e^{-\frac{t_{1}+t_{2}}{\tau_{B}}}\nonumber\\
&\times&\left<f^{a}\left(t_{1}\right)f^{b}\left(t_{2}\right)\right>.\nonumber\\
\end{eqnarray}
Next,  we use the fluctuation-dissipation theorem  (\ref{locfluc-diss}). Thus the integration over variable $t_{2}$ leads to the following  result
\begin{eqnarray}
\int_{0}^{t^{\prime}}dt_{2}e^{t_{2}/\tau_{B}}\delta\left(t_{2}-t_{1}\right)=\theta\left(t^{\prime}-t_{1}\right)e^{t_{1}/\tau_{B}},
\end{eqnarray}
where $\theta(x)$ is the Heaviside step-function. The remaining integral over $t_{1}$ can be done for two cases $t^{\prime}>t$ and $t^{\prime}<t$. If $t^{\prime}>t$ then $t^{\prime}>t_{1}$ for all $t_{1}\in\left[0, t\right]$, therefore $\theta(t^{\prime}-t_{1})=1$. Now, if $t^{\prime}<t$ then the integration for $t_{1}$ can be splitted in two parts
\begin{eqnarray}
\int_{0}^{t} dt_{1}\theta\left(t^{\prime}-t_{1}\right)e^{2t_{1}/\tau_{B}}&=&\int_{0}^{t^{\prime}} dt_{1}\theta\left(t^{\prime}-t_{1}\right)e^{2t_{1}/\tau_{B}}\nonumber\\&+&\int_{t^{\prime}}^{t} dt_{1}\theta\left(t^{\prime}-t_{1}\right)e^{2t_{1}/\tau_{B}}.\nonumber\\
\end{eqnarray}
In the first integral $t^{\prime}>t_{1}$, since $t_{1}\in\left[0, t^{\prime}\right]$. Then for this integral  $\theta\left(t^{\prime}-t_{1}\right)=1$. For the second integral, we have $t^{\prime}<t_{1}$, since $t_{1}\in\left[t^{\prime}, t\right]$. Therefore $\theta\left(t^{\prime}-t_{1}\right)=0$. Now, joining these results and performing the elementary integrals we reproduce equation (\ref{pp2}).

\subsection{Calculation of $\mathcal{J}$ function}

The determination of $\mathcal{J}\left(t/\tau_{B}\right)\equiv J(t)/\left(\tau_{B}D_{0}\right)^{2}$ can be obtained from the calculation of
\begin{eqnarray}
J(t)&=&\frac{1}{M^{4}}\int_{0}^{t}dt_{1}\int_{0}^{t}dt_{2}\int_{0}^{t_{2}}dt_{3}\int_{0}^{t_{3}}dt_{4}e^{-\frac{1}{\tau_{B}}(t_{2}-t_{3})}\nonumber\\
&&\times\left(G(t_{1},t_{4})G(t_{3},t_{3})-G(t_{1},t_{3})G(t_{4},t_{3})\right).\nonumber\\
\label{Jfunction}
\end{eqnarray}
We should remark that the integral $I\left(t,\tau^{\prime}\right)=\int^{t}_{0}d\tau G(\tau,\tau^{\prime})$ appears in various places in the multiple integral (\ref{Jfunction}).  Thus, the function (\ref{Jfunction}) can be written as follows
\begin{eqnarray}
J(t)&=&\frac{1}{M^{4}}\int_{0}^{t}dt_{2}\int_{0}^{t_{2}}dt_{3}e^{-\frac{1}{\tau_{B}}(t_{2}-t_{3})}\nonumber\\
&&\times\left(G(t_{3},t_{3})\int_{0}^{t_{3}}dt_{4}I(t,t_{4})-I(t,t_{3})I(t_{3},t_{3})\right).\nonumber\\
\label{Jfunction-r}
\end{eqnarray}
The advantage to write $J(t)$ in terms of $I(t, \tau^{\prime})$ is that $\tau^{\prime}\leq t$. For the calculation of the function $I(t, \tau^{\prime})$ it is convenient to use the following equivalent expression for the Green function 
\begin{eqnarray}
G\left(t,t^{\prime}\right)&=&\tau_{B}\Omega\left[e^{-\frac{t}{\tau_{B}}}\theta\left(t-t^{\prime}\right)\sinh \frac{t^{\prime}}{\tau_{B}}\right.\nonumber\\&+&\left.e^{-\frac{t^{\prime}}{\tau_{B}}}\theta\left(t^{\prime}-t\right)\sinh \frac{t}{\tau_{B}}\right].
\end{eqnarray}
Performing its integral we obtain
\begin{eqnarray}
I(t, \tau^{\prime})&=&\frac{\tau_{B}^2\Omega}{2}e^{-\frac{1}{\tau_{B}}\left(t+\tau^{\prime}\right)}
\left(1-e^{\frac{\tau^{\prime}}{\tau_{B}}}\right)\nonumber\\&\times&\left(1-2e^{\frac{t}{\tau_{B}}}+e^{\frac{\tau^{\prime}}{\tau_{B}}}\right).\nonumber\\
\end{eqnarray}
Now, we carry out the elementary integrations involved in $(\ref{Jfunction-r})$. We then get the following expression
{\small\begin{eqnarray}
J(t)&=&\frac{1}{6}\left(\tau_{B}D\right)^{2}\left\{e^{-\frac{t}{\tau_{B}}}\left[8+e^{-3\frac{t}{\tau_{B}}}-8e^{-2\frac{t}{\tau_{B}}}+36e^{-\frac{t}{\tau_{B}}}\right.\right.\nonumber\\&+&\left.\left.48\frac{t}{\tau_{B}}+e^{\frac{t}{\tau_{B}}}\left(-37+12\frac{t}{\tau_{B}}\right)\right]+
e^{-4\frac{t}{\tau_{B}}}\left[1+e^{\frac{t}{\tau_{B}}}\right.\right.\nonumber\\&\times&\left.\left.\left(-8+e^{\frac{t}{\tau_{B}}}\left(15-40e^{\frac{t}{\tau_{B}}}-6\frac{t}{\tau_{B}} +  e^{2\frac{t}{\tau_{B}}}\left(32\right.\right.\right.\right.\right.\nonumber\\&+&\left.\left.\left.\left.\left.6\frac{t}{\tau_{B}}\left(\frac{t}{\tau_{B}}-4\right) \right)\right)\right)\right]\right\}.\nonumber\\
\end{eqnarray}}

\section{Expectation values for Brownian motion over  $S^{2}$}

\noindent  The expectation values for $\mathcal{O}=\mathcal{O}(\theta,\varphi)$ can be calculated from 
\begin{eqnarray}
\label{eqomega}
\left<\mathcal{O}\left(\theta,\varphi\right)\right>=\frac{R}{2}\sum^{\infty}_{\ell=0}\left(2\ell+1\right)g_{\mathcal{O}}\left(\ell\right)e^{-D\ell\left(\ell+1\right)t/R^2}, 
\label{expectation}
\end{eqnarray}
where
\begin{eqnarray}
\label{eggo}
g_{\mathcal{O}}\left(\ell\right)=\int^{\pi}_{0}\int^{2\pi}_{0}d\theta d\varphi \sin\theta~\mathcal{O}\left(\theta,\varphi\right)P_{\ell}\left(\cos\theta\right).
\label{g-function}
\end{eqnarray}
Equation (\ref{eggo}) depends explicitly on the chosen form of $\mathcal{O}$. In general, equation (\ref{eqomega}) cannot be written in a closed form and it must be studied numerically. In particular, we discuss here the mean values of the functions $\mathcal{O}=s=R\theta$,  and $\mathcal{O}=s^2$. In order to have a more manageable form for these expectation values  we use the following identity
\begin{eqnarray}
P_{\ell}\left(\cos\theta\right)=\left(-1\right)^{\ell}\sum^{\ell}_{k=0}\left(\begin{array}{c}
-\frac{1}{2}\\
\ell
\end{array}\right)
\left(\begin{array}{c}
-\frac{1}{2}\\
\ell-k
\end{array}\right)\cos\left[\left(\ell-2k\right)\theta\right].\nonumber\\
\end{eqnarray}
Now, in order to obtain (\ref{g-function}) we perform the integration for even and odd vaÂues of $\ell$. After performing the ele\-men\-ta\-ry integrations, we obtain the following results.  For $s=R\theta$, $g_{s}\left(\ell\right)$ is zero for even values of $\ell$, and  for odd values of $\ell$ it takes the form
\begin{eqnarray}
g_{s}\left(2p+1\right)&=&\frac{\pi}{2}\left(\begin{array}{c}
-\frac{1}{2}\\
p
\end{array}\right)\left(\begin{array}{c}
-\frac{1}{2}\\
p+1
\end{array}\right)\nonumber\\
&-&\pi\sum^{2p+1}_{k=0} \frac{\left(\begin{array}{c}
-\frac{1}{2}\\
k
\end{array}\right)\left(\begin{array}{c}
-\frac{1}{2}\\
2p+1-k
\end{array}\right)}{\left(2\left(p-k\right)+1\right)^2-1},\nonumber\\
\end{eqnarray}
where the last sum does not take the values $k=p$ and $k=p+1$. For $s^{2}=R^{2}\theta^{2}$, it is not difficult to show the identity $g_{s^{2}}\left(2p+1\right)=\pi g_{s}\left(2p+1\right)$ for odd values of $\ell$. However, for even values of $\ell$ we find
\begin{eqnarray}
g_{s^{2}}\left(2p\right)=
\sum^{2p}_{k=0}\left(\begin{array}{c}
-\frac{1}{2}\\
k
\end{array}\right)\left(\begin{array}{c}
-\frac{1}{2}\\
2p+1-k
\end{array}\right)H\left(2\left(p-k\right)\right),\nonumber\\
\end{eqnarray}
where $H$ is a function defined as 
\begin{equation*}
H\left(z\right)\equiv\frac{12 z^2+4-\pi^2\left(z^2-1\right)^2}{\left(z^2-1\right)^{3}},
\end{equation*}
and  
\begin{equation*}
\left(\begin{array}{c}
x\\
n
\end{array}\right)=x\left(x-1\right)\left(x-2\right)\cdots\left(x-n+1\right)/n!
\end{equation*}
is the binomial coefficient \cite{Gradshteyn}.

\section{Partition function on $\mathbb{M}$, Error identity and pseudo-corrective force}

\noindent{\it Partition function.} The Hamiltonian for a free particle of mass $M$ on a  $d$-dimensional Riemannian manifold $\mathbb{M}$ is given by  $\mathcal{H}\left(p,q\right)=\frac{1}{2M}g^{ab}p_{a}p_{b}$, where $g_{ab}$ is the metric tensor of $\mathbb{M}$. The single partition function associated to this Hamiltonian is given by
\begin{eqnarray}
\mathbb{Z}_{\mathbb{M}}&=&\int\prod^{d}_{a=1}\left(\frac{dp_{a}dq^{a}}{2\pi\hbar}\right)\exp\left(-\beta\mathcal{H}\left(p,q\right)\right)\nonumber\\
&=&\frac{1}{\lambda^{d}\left(T\right)}\int d^{d}q\sqrt{\det g}=v/\lambda^{d}\left(T\right)\nonumber,
\end{eqnarray}
where $v$ is the volume of $\mathbb{M}$, assumed it is compact. 
\vskip0.5em
\noindent{\it The error function}. The exact value of the integral in equation (\ref{Integral}) is, 
\begin{eqnarray}
I\left(x_{0}\right)=\int^{\infty}_{-x_{0}}dy\left(y+x_{0}\right)^2\exp\left(-y^{2}\right),
\end{eqnarray}
\begin{eqnarray}
I\left(x_{0}\right)=\frac{x_{0}}{2}\exp\left(-x^{2}_{0}\right)+\frac{\sqrt{\pi}}{4}\left(1+2x^2_{0}\right)\left(1+\erf\left(x_{0}\right)\right).\nonumber\\
\end{eqnarray}
Using the asymptotic behaviour of the Error function \cite{Gradshteyn} for large values of $x_{0}$, we have the following asymptitic behavior
\begin{eqnarray}
\frac{I\left(x_{0}\right)}{\sqrt{\pi}x^{2}_{0}}\approx 1+\frac{1}{2x^{2}_{0}}+O\left(\exp\left(-x^2_{0}\right)/x_{0}\right).
\end{eqnarray}

\vskip0.5em
\noindent{\it Pseudo-corrective force}. According to E. J. Hinch \cite{Hinch} in order to convert the Brownian motion of a rigid system to that of an equivalent very stiff system, we should add the following pseudo-corrective force to the equations of motion,
\begin{eqnarray}
F^{\left(c\right)}_{i}=-\frac{\partial}{\partial x_{i}}\left(kT\ln\sqrt{\det}\right),
\label{fuerzacorrectiva}
\end{eqnarray}
where $\det=\det\sum_{i}m^{-1}_{i}\frac{\partial g^{a}}{\partial x_{i}}\cdot\frac{\partial g^{b}}{\partial x_{i}}$, and $g^{a}$ is the constraint function of the system.  For the corresponding conditions, the determinant is reduced to $\det=\frac{1}{M}\left|\nabla\Phi\right|^{2}$. For this case, the corrective force becomes
\begin{eqnarray}
F^{\left(c\right)i}=-kT n_{j}~G^{ji},
\label{fuerzacorrectiva1}
\end{eqnarray}
where ${\bf n}$ is the normal vector of the hypersurface and $G_{ij}$ is the $G$-matrix defined above. For the spherical case, the constraint function is given by $\Phi\left({\bf x}\right)=\frac{{\bf x}^{2}}{R^2}-1$, where $R$ is the radius of the sphere. In this case the pseudo-potential force is ${\bf F}^{\left(c\right)}=-\frac{kT}{R}{\bf n}$. A similar result occurs for the circle.

\begin{acknowledgments}
Financial support by PIFI-2011, PIEC, PROMEP (1035/08/3291), and CONACyT (through grants 61418/2007, 102339/2008, 60595 and Red Tem\'atica de la Materia Condensada Blanda) is kindly acknowledged.
\end{acknowledgments}


\begin{thebibliography}{99}
\bibitem{Soft}  E. Frey and K. Kroy, Annalen der Physik, {\bf 14}, 2050 (2005).
\bibitem{quark-diffusion} Benjamin Svetitsky, Phys. Rev. D {\bf  37},  2484 (1988).

\bibitem{Hu} Bei Lok Hu and Enric Verdaguer, Living Rev. Relativity, {\bf 11}, 3 (2008).

\bibitem{biophysics} Nina Malchus and Matthias Weiss, Biophysical Journal, {\bf 99}, 1321 (2010); Matthias Weiss, Hitoshi Hashimoto, and Tommy Nilsson, Biophysical Journal {\bf 84}, 4043 (2003); Valerii M. Sukhorukov, J$\ddot{u}$rgen Bereiter-Hahn, PLoS {\bf 4} e4604 (2009) .

\bibitem{Alberts} Bruce Alberts, Alexander Johnson, Julian Lewis, Martin Raff, Keith Roberts, and Peter Walter, {\it Molecular Biology of the Cell}, 4th edition {\it  Garland Science} (2002).

\bibitem{Naji} Ali Naji, and Franck L. H. Brown,  J. Chem. Phys. {\bf 126}, 235103 (2007)
\bibitem{fidel1} F. C\'ordoba-Vald\'es, C. Fleck and R. Casta\~neda-Priego, Rev. Mex. Fis. {\bf 53},  475 (2007); F. C\'ordoba-Vald\'es, C. Fleck, J. Timmer and R. Casta\~neda-Priego, {\it submitted}.
\bibitem{fidel2} Ali Naji, Paul J. Atzberger, and Frank L. H. Brown, Phys. Rev. Lett.  {\bf 102 },138102 (2009).  
\bibitem{Reister1} Ellen Reister-Gottfried, Stefan M. Leitenberger, and Udo Seifert, Phys. Rev. E {\bf 75}, 011908 (2007); Ellen Reister-Gottfried, Stefan M. Leitenberger, and Udo Seifert, Phys. Rev. E {\bf 81}, 031903 (2010).


\bibitem{faraudo02} Faraudo J, J. Chem. Phys. {\bf 116}, 5831 (2002).

\bibitem{Gustafsson}  S. Gustafsson and B. Halle, J. Chem. Phys.  {\bf 106}, 1880 (1997).
\bibitem{Reister} Reister E and Seifert U, Europhys Lett. {\bf 71}, 859 (2005)

\bibitem{Gov} N. S. Gov, Phys. Rev. E {\bf 73}, 041918 (2006).
\bibitem{naohisa} Naohisa Ogawa, Phys. Rev. E {\bf 81}, 061113 (2010). 

\bibitem{Aizenbaud} Boris M. Aizenbud and Nahum D. Gershon, Biophys. J. {\bf 38}, 287 (1982).
\bibitem{H}  D. Anderson and H. Wennerstr$\ddot{o}$m, J. Phys. Chem. {\bf 94}, p. 8683 (1990).
\bibitem{K} J. Balakrishnan, Phys. Rev. E {\bf 61}, 4648 (2000)
\bibitem{Holyst} R. Holyst, D. Plewczynski, A. Aksimentiev, and K. Burdzy, Phys. Rev. E {\bf 60}, 302 (1999).
\bibitem{Christensen} Micheal Christensen, Journal of Computational Physics {\bf 201}, 421-438 (2004).
\bibitem{Tomoyoshi} Tomoyoshi Yoshigaki, Phys. Rev. E {\bf 75}, 041901 (2007).


\bibitem{Castro} Pavel Castro-Villarreal,  J. Stat. Mech.  P08006 (2010).

\bibitem{Kampen} N. G. van Kampen, J. Stat. Phys. {\bf 44}, Nos. 1/2 (1986).

\bibitem{Fixman} Marshall Fixman, J. Chem. Phys. {\bf 69}, 1527 (1978).
\bibitem{Rallison} J. M. Rallison, J. Fluid. Mech. {\bf 93}, 251 (1979).
\bibitem{Hinch} E. J. Hinch, J. Fluid. Mech. {\bf 271}, 219 (1993).
\bibitem{Ottinger}  Hans Christian $\ddot{\rm O}$ttinger, Phys. Rev. E {\bf 50} 2696 (1994). Hans Christian $\ddot{\rm O}$ttinger, {\it Stochastic processes in polymeric fluids}. Ed. Springer, (1996).
\bibitem{Morse} David C. Morse, Adv. Chem. Phys. {\bf 128}, 65 (2004).

\bibitem{KampenandLodder} N. G. van Kampen and J. J. Lodder, Am. J. Phys {\bf 52}, 419 (1984).

\bibitem{Kleinert} H. Kleinert and S. V. Shabanov, J. Phys. A: Math. Gen. {\bf 31}, 7005-7009 (1998).
\bibitem{Baush} Richard Baush, Rudi Schmitz, and \L ukasz A. Turski,  Z. Phys. B. {\bf 97}, 171 (1995); Richard Baush, Rudi Schmitz, and \L ukasz A. Turski {\bf 73}, 2382 (1994).
\bibitem{Lingand Chen} Lingang Chen and Micheal W. Deem, Phys. Rev. E {\bf 68}, 021107 (2003).

\bibitem{Smerlak} Matteo Smerlak, New Journal of Phys. {\bf 14},  023019 (2012); Matteo Smerlak, Phys. Rev. E {\bf 85}, 041134 (2012).
\bibitem{Dhont} J. K. G. Dhont, {\it An introdution to dynamics of colloids}, Ed. Elsevier, (1996).

\bibitem{Polettini} Matteo Polettini, ArXive: 1206.2798v2 (2012).

\bibitem{Grygorian} Grugor'yan A,  London Mathematical Society Lecture Note Series {\bf 273}, 140 (1999).

\bibitem{Kramers} H. A. Kramers, J. Chem. Phys. {\bf 14} 415 (1946).

\bibitem{Ermak} Donald L. Ermak and J. A. McCammon, J. Chem. Phys. {\bf 69}, 1352 (1978).

\bibitem{Zinn-Justin} Jean Zinn-Justin, {\it Quantum Field Theory and Critical Phenomena}, 3rd Ed. Oxford (1995).

\bibitem{Spivak} M. Spivak, {\it A Comprehensive Introduction to Differential Geometry} Vol. 3,  3rd Ed. 1999.

\bibitem{muller} M$\ddot{u}$ller U, Schubert C, and van de Ven A E M, Gen. Rel. Grav. {\bf 31}, 1759 (1999).

\bibitem{Jean} Jean-Michel Caillol, Phys. A: Math. Gen. {\bf 37}, 3077-3083 (2004). 	

\bibitem{Vassilevich} D. V. Vassilevich, Phys. Rep. {\bf 388}, 279 (2003).
\bibitem{chavel} I. Chavel, {\it Eigenvalues in Riemannian geometries} (Academic Press, 1984).

\bibitem{Chava2010} S. Herrera-Velarde, A. Zamudio-Ojeda and R. Casta\~eda-Priego, J. Chem. Phys.  {\bf 133}, 114912 (2010).


\bibitem{muthu98} Radu P. Mondescu and M. Muthukumar,  Phys. Rev E {\bf 57}, 4411 (1998).

\bibitem{ghosh} Abhijit Ghosh1, Joseph Samuel2 and Supurna Sinha2, EPL {\bf 98} (2012).


\bibitem{DifRot} Th. Kirchhoff, H. L\"owen, and R. Klein,  Phys. Rev. E {\bf 53}, 5011 (1996).


\bibitem{Duenweg} Tri T. Pham, Ulf D. Schiller, J. Ravi Prakash, and B. D\"unweg, J. Chem. Phys.  {\bf 131}, 164114 (2009).

\bibitem{Grassia} P. S. Grassia, E. J. Hinch and L.C. Nitcshe, J. Fluid. Mech. {\bf 282} 373 (1995).


\bibitem{GLEMedina} M. Medina-Noyola,  Faraday Discuss. Chem. Soc. {\bf 83},  21 (1987).




\bibitem{Gradshteyn} Gradshteyn and Ryzhik's, {\it Table of Integrals, Series, and Products} {\it Academic press} Seventh edition (Feb 2007).











\end{thebibliography}
\end{document}